\documentclass[journal]{IEEEtran}

\usepackage{cite}
\usepackage{amsmath,amssymb,amsfonts}
\usepackage{algorithmic}
\usepackage{graphicx}
\usepackage{textcomp}
\usepackage{xcolor}
\usepackage{cuted}
\usepackage{epstopdf}
\usepackage{stfloats}
\usepackage{algorithm}
\usepackage{enumerate}
\usepackage{caption}
\usepackage{comment}

\usepackage[caption=false,font=footnotesize,labelfont=rm,textfont=rm]{subfig}

\usepackage{svg}

\begin{document}

\title{Movable Intelligent Surface (MIS) for Wireless Communications: Architecture, Modeling, Algorithm, and Prototyping}
\author{Ziyuan~Zheng, Qingqing~Wu, Wen~Chen, Xiangming~Wu and
        Weiren~Zhu     \vspace{-12pt}
\thanks{The authors are with the Department of Electronic Engineering, Shanghai Jiao Tong University, 200240, China (e-mail: \{zhengziyuan2024, qingqingwu, wenchen, wxm0805, weiren.zhu\}@sjtu.edu.cn).}
}

\markboth{}%
{Shell \MakeLowercase{\textit{et al.}}: Bare Demo of IEEEtran.cls for IEEE Journals}

\maketitle

\begin{abstract}
Reconfigurable intelligent surfaces (RISs) enhance wireless systems by reshaping propagation environments. However, dynamic metasurfaces (MSs) with numerous phase-shift elements may incur undesired control overhead and hardware costs. In contrast, static MSs (SMSs), configured with static phase shifts that are pre-designed for specific communication demands, offer a cost-effective alternative by eliminating agile element-wise tuning. Nevertheless, SMSs typically support only a single beam pattern, limiting flexibility in dynamic and multi-user scenarios.
In this paper, we propose a novel Movable Intelligent Surface (MIS) technology that enables dynamic beamforming while maintaining static phase shifts. Specifically, we design a MIS architecture comprising two closely stacked transmissive MSs: a larger fixed-position MS 1 and a smaller movable MS 2. By differentially shifting MS 2's position relative to MS 1, the MIS synthesizes distinct desired beam patterns, overcoming the SMSs' single-pattern limitation.
Then, we model the interaction between MS 2 and MS 1 using binary selection matrices and padding vectors, which allow us to formulate a new optimization problem that jointly designs the MIS phase shifts and selects shifting positions for worst-case signal-to-noise ratio (SNR) maximization. This position selection, equal to beam pattern scheduling, offers a new degree of freedom for RIS-aided systems. To solve the intractable problem, we develop an efficient algorithm that handles unit-modulus and binary constraints and employs manifold optimization methods. 
Finally, extensive validation results are provided, including both experimental and numerical analysis. We first implement a MIS prototype and perform proof-of-concept experiments, demonstrating the MIS's ability to synthesize desired beam patterns that achieve one-dimensional beam steering. Numerical results further show that by introducing a movable MS 2 with a few elements, MIS effectively offers beamforming flexibility for significantly improved performance compared to SMSs. We also draw insights into the optimal MIS configuration and element allocation strategy.
\end{abstract}


%
\IEEEpeerreviewmaketitle

\section{Introduction}
As wireless communication systems evolve toward the sixth generation (6G), the demand for higher data rates, improved coverage, and increased spectral efficiency continues to grow [1]. Reconfigurable Intelligent Surface (RIS) has emerged as a promising technology to meet these demands by reconfiguring the wireless propagation environment [2]. By strategically adjusting the phase shifts of massive reflecting or transmissive elements, RIS enables passive beamforming in cooperation with transmitters. Unlike traditional active relay systems, RIS operates without requiring power amplification or radio-frequency (RF) chains, offering low cost, low power consumption, and no self-interference in full-duplex mode [3]. These appealing advantages position RIS as a powerful tool to enhance the performance of existing wireless systems.

In recent years, the concept of RIS has gained widespread attention and recognition, and extensive research has explored various aspects of RIS technology. Studies have investigated deployment strategies at the link and network levels [4], efficient channel estimation techniques with different hardware capabilities [5], and sophisticated beamforming designs for diverse wireless applications [6]. RIS shows particular promise for deployment in urban and indoor environments, extending coverage and bypassing obstructions by providing virtual line-of-sight (LoS) links, effectively reducing the power consumption of active relays or repeaters [7]. Furthermore, RIS can improve channel rank, enhance spatial multiplexing, and mitigate multi-user interference, leading to spectral and energy efficiency gains in multiple-input multiple-output (MIMO) communication systems, thus alleviating the burden of massive active antenna arrays [8], [9]. Additionally, integrating RIS into large-scale cell-free or distributed MIMO networks has attracted attention, offering a cost-effective and energy-efficient solution for widespread coverage without needing dense base station (BS) deployments [10], [11].

Building upon the fundamental intelligent reflecting surface (IRS) architecture [12], several multifunctional RIS architectures have been proposed, including intelligent transmissive surfaces [13], simultaneous transmission and reflection RIS (STAR-RIS) [14], and intelligent omni-surfaces (IOS) [15]. Basically, these architectures typically feature a single-layer element-wise tunable metasurface (MS) with a diagonal phase shift matrix. Recent developments have introduced advanced RIS designs, such as beyond-diagonal RIS (BD-RIS) and stacked intelligent MS (SIM) [16], [17]. Specifically, BD-RIS characterizes a RIS as multiple antennas connected to a group-connected multi-port reconfigurable impedance network, enabling more flexible passive beamforming through partially or fully cell-wise connected modes; SIM incorporates multiple stacked MS layers to perform advanced signal processing and beamforming refinement in a native electromagnetic wave regime, emphasizing its specific layer-by-layer hardware structure. While these innovations extend the boundaries of RIS performance, they also increase certain hardware costs and operational complexity from a technical practicality standpoint.

Despite the promising theoretical potential of these advanced RIS architectures, it is crucial to revisit the original motivation behind IRS/RIS technology: to provide a cost-effective, energy-efficient, and low-complexity solution to improve wireless communication [2]. However, increasingly sophisticated hardware implementations of advanced RIS architectures may be overly ambitious, relying on further developments in electronics and materials science [6]. Moreover, while classic dynamic RIS yields impressive beamforming flexibility and performance gains, achieving this requires a rapid, frequent, and element-wise reconfiguration in a large number of adjustable phase shifters [18]. This idealized reconfiguration ability may place stringent and even impractical demands on MSs. Meanwhile, accurate adjustment of dynamic RIS incurs significant signaling and control overhead, further increasing operational burdens on smart controllers and power consumption of electronic components [19]. As a result, these challenges complicate the practical application and widespread deployment of dynamic RIS systems, presenting a performance-cost trade-off that needs to be balanced.

In contrast to the pursuit of multifunctionality and sophistication, static MS (SMS) (or static surface), configured with a set of pre-designed phase shifts tailored to specific communication demands, offers a promising solution to practical challenges related to hardware cost, power consumption, and control overhead by avoiding the need for agile element-wise phase tuning [20]. For example, with a beam-flattening design, an SMS can be deployed to assist an aerial BS in extending the communication coverage to a specific terrestrial area of interest [21]. While SMS presents a low-cost alternative that preserves key performance benefits, such as virtual LoS link establishment and coverage extension, it faces limitations due to reduced beamforming flexibility. Specifically, a single set of static phase shifts in an SMS typically forms only one beam pattern, restricting beam positions and coverage directions. Thus, SMSs are less adaptable in dynamic environments or multi-user scenarios. Although the integration of distributed MIMO principles can improve performance by exploiting spatial diversity through access point selection [22], the inability of SMSs to dynamically generate multiple beam patterns remains an inherent limitation of this architecture.

Parallel to RIS and MS developments, movable antenna (MA) technology [23], also known as fluid antenna system (FAS) [24], has recently been introduced to enhance wireless communications. These technologies leverage antenna repositioning or port selection to adapt to wireless environments and reshape channel conditions. By merely adjusting MA's position in communication transmitters or receivers, it is possible to dynamically steer the beam or alter the coverage area, achieving performance gains with minimal RF chains [25]. 

Inspired by the above advances, this paper introduces a novel \textbf{Movable Intelligent Surface (MIS)} technology that \textit{enables dynamic beamforming while maintaining static phase shifts.}
The main contributions of this paper are as follows.
\begin{enumerate}
\item We propose a novel MIS architecture comprising two closely stacked transmissive MSs: a larger fixed-position MS 1 and a smaller movable MS 2. Building on this architecture, we propose the differential position shifting mechanism that moves the position of MS 2 as a whole in discrete units relative to and within MS 1. In this way, the MIS synthesizes desired distinct beam patterns through the variation of superimposed phase shifts, eliminating the need for element-wise phase tuning.
\item We model the interaction between MS 2 and MS 1 by binary selection matrices and padding vectors and characterize the signal model of the MIS communication system. Following the modeling, we formulate a new optimization problem that jointly designs MIS phase shifts and selects the shifting positions to maximize the worst-case signal-to-noise ratio (SNR) in a target coverage area. This position selection, equal to beam pattern scheduling, offers a new degree of freedom for RIS-aided systems.
\item We develop an efficient algorithm to solve the intractable mixed-integer non-convex non-smooth optimization problem. Specifically, we first address the non-smooth max-min function with smoothing techniques and relax the binary nature of scheduling variables to form a multinomial manifold constraint. Then, we employ manifold optimization techniques over a constructed product manifold, enabling simultaneous updates of all optimization variables within an iterative framework.
\item We implement a MIS prototype and perform proof-of-concept experiments to validate the ability to synthesize desired beam patterns through differential position shifting of MSs. The fabricated transmissive MIS achieves one-dimensional (1D) beam steering with a steering angle of ±45° at 12.2 GHz and maintains a gain fluctuation of less than -3 dB. Subsequently, we present numerical results to evaluate various MIS configurations. Notably, introducing a movable MS 2 with a few elements significantly improves the worst-case SNR compared to a single-layer SMS. Furthermore, with a fixed total number of MIS elements, allocating a moderate number of elements to MS 2 achieves optimal gains, providing insight into the allocation of MIS elements.
\end{enumerate}

The remainder of this paper is organized as follows. Section II designs a specific MIS architecture. Section III presents the MIS communication model and formulates the optimization problem. Section IV proposes the algorithm based on manifold optimization. Section V provides the experimental and numerical results. Finally, Section VI concludes the paper.

\section{MIS Architecture Design}
\begin{figure}[t]
    \centering
    \hspace{-5pt}\includegraphics[width=3.4in]{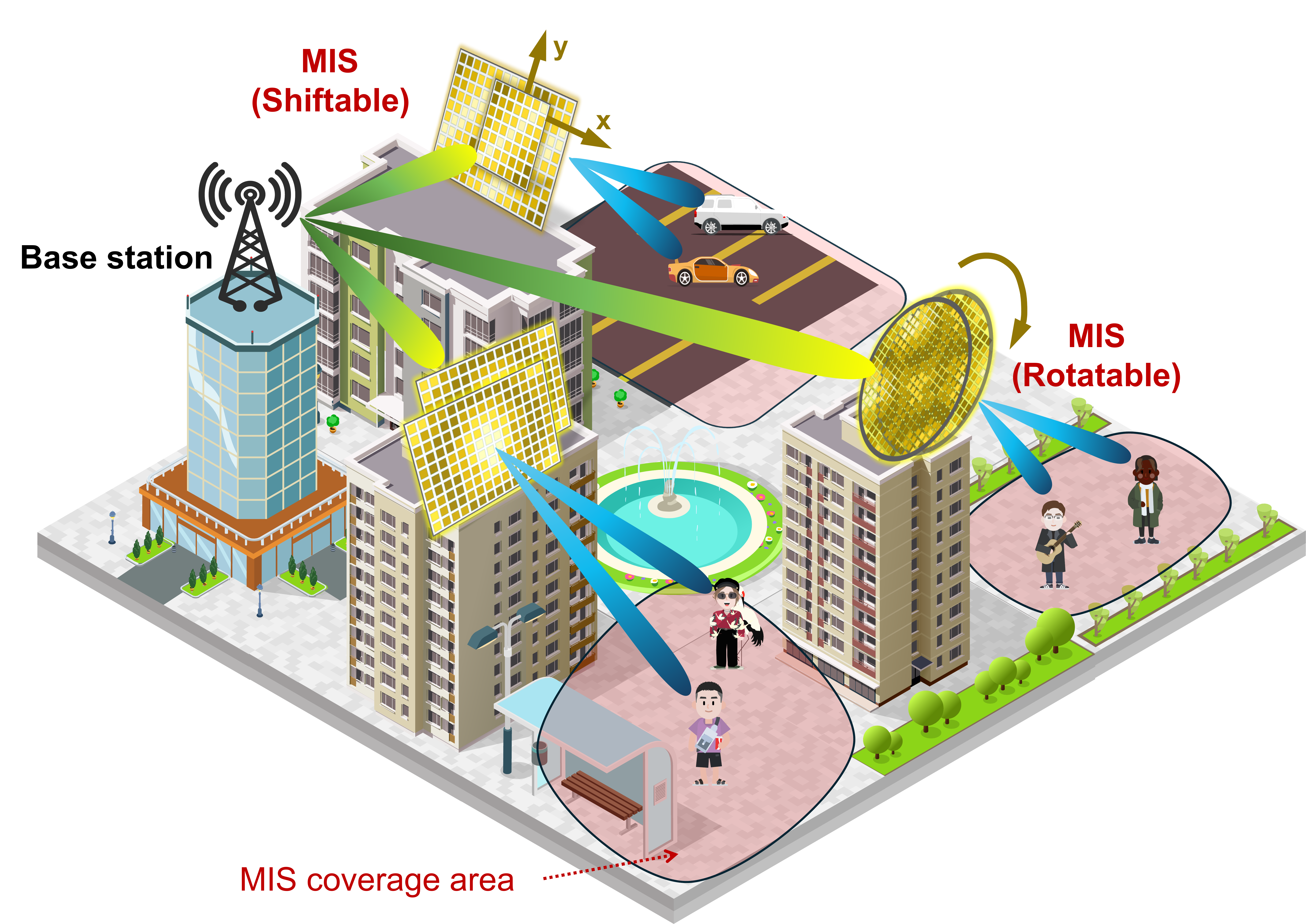}
    \captionsetup{font=small}
    \caption{Illustration of MIS-enabled wireless communication systems.} 
    \vspace{-15pt}
    \label{fig:system_model}
\end{figure}
As shown in Fig. 1, we consider a MIS-aided wireless relaying system, where transmissive MISs are deployed to assist an $L$-antenna BS in extending its communication coverage to target areas. We assume that the direct link from the BS to the target area is obstructed.\footnote{The work in this paper can also be extended to the case with a direct link.} 
Specifically, by establishing a MIS-enabled virtual LoS link, the signal can be refracted from the BS to the desired locations or directions for users in a target coverage area.\footnote{To illustrate, consider a 1D beam scanning scenario where the base station (BS) utilizes the MIS to serve $K$ users uniformly distributed across azimuth angles $\nu_k \in \left[\nu_{\text{ini}}, \nu_{\text{fin}}\right]$ along one side of the MIS, all at a fixed elevation angle.} In this section, we introduce a novel MIS architecture with two mutually shifted MSs and then elaborate on its beam steering mechanism, which does not require element-wise phase adjustment.

\subsection{MIS Architecture and Configuration}
The MIS comprises two transmissive MSs, namely MS 1 and MS 2, of different sizes, employing pre-designed static phase shifts on both MSs. These two MSs are closely stacked, and MS 2, which is smaller, can slide onto and within MS 1 on a grid in units of transmissive elements. The detailed configuration of the MIS is described below. 
\begin{itemize}
    \item \textbf{MS 1}, the larger fixed-position MS, consists of $M_r$ rows and $M_c$ columns transmissive elements, that is, totaling $M = M_r \times M_c$ elements. The elements are spaced $d_{\text{MIS}} = \lambda/2$ apart, where $\lambda$ denotes the carrier wavelength. Each element $m\in \mathcal{M}=\{1, 2, \dots, M\}$ can be indexed according to its row and column positions with $m = (m_r-1)\times M_c +m_c$ where $m_r \in \{1, 2, \dots, M_r\}$ and $m_c \in \{1, 2, \dots, M_c\}$.   
    \item \textbf{MS 2}, the smaller movable MS, closely stacked with MS 1, consists of transmissive elements of $N_r$ rows and $N_c$ columns, that is, totaling $N = N_r \times N_c$ elements. Similarly to the definition of MS 1, the element spacing of MS 2 is also $d_{\text{MIS}}$. The elements are indexed according to their row and column positions by $n = (n_r-1)\times N_c +n_c \in \mathcal{N} = \{1, 2, \dots, N\}$. Conceptually speaking, MS 2 can adjust its position along MS 1's columns or rows constrained within the two-dimensional (2D) boundaries of MS 1.\footnote{In practice, slide rails and motors can implement this movement in MISs.} Through movements relative to MS 1, MS 2 will overlap with different subsets of elements in MS 1. 
\end{itemize}
The phase shift matrices for MS 1 and MS 2 are denoted by $\boldsymbol{\varPhi} = \mathrm{diag}(\boldsymbol{\phi})\in \mathbb{C}^{M\times M}$ and $\boldsymbol{\varTheta} = \mathrm{diag}(\boldsymbol{\theta})\in \mathbb{C}^{N\times N}$, where $\boldsymbol{\phi }=\left[ e^{\phi _1},\dots,e^{\phi _M} \right] ^T$ and $\boldsymbol{\theta }=\left[ e^{\theta _1},\dots,e^{\theta _N} \right] ^T$ are the corresponding transmission phase shift vectors. Each phase shift $\{\phi_m, m\in \mathcal{M}\}$ and $\{\theta_n, n\in \mathcal{N}\}$ is in the range $[0, 2\pi]$.\footnote{Here, we assume continuous phase shifts. Discrete phase shifts can be considered in future work to account for more practical hardware limitations.}

It is worth mentioning that, unlike traditional RISs that dynamically adjust phase shifts, our proposed MIS employs static phase elements in both MS 1 and MS 2. These elements are pre-designed to achieve dynamic beamforming or beam steering, significantly reducing hardware costs associated with phase-adjustable RIS components and sophisticated RIS controllers. Furthermore, the MIS architecture enables fully passive implementation, avoids coupling effects, and is better suited for high-frequency wireless systems. 

\label{subsec:Proposed_Architecture_Beam_Steering}
\setlength{\abovecaptionskip}{6pt}
\begin{figure}[t]
    \subfloat[Synthesizing Beam pattern 1 with MS 2 at initial position.]{
        \includegraphics[width=3.32in]{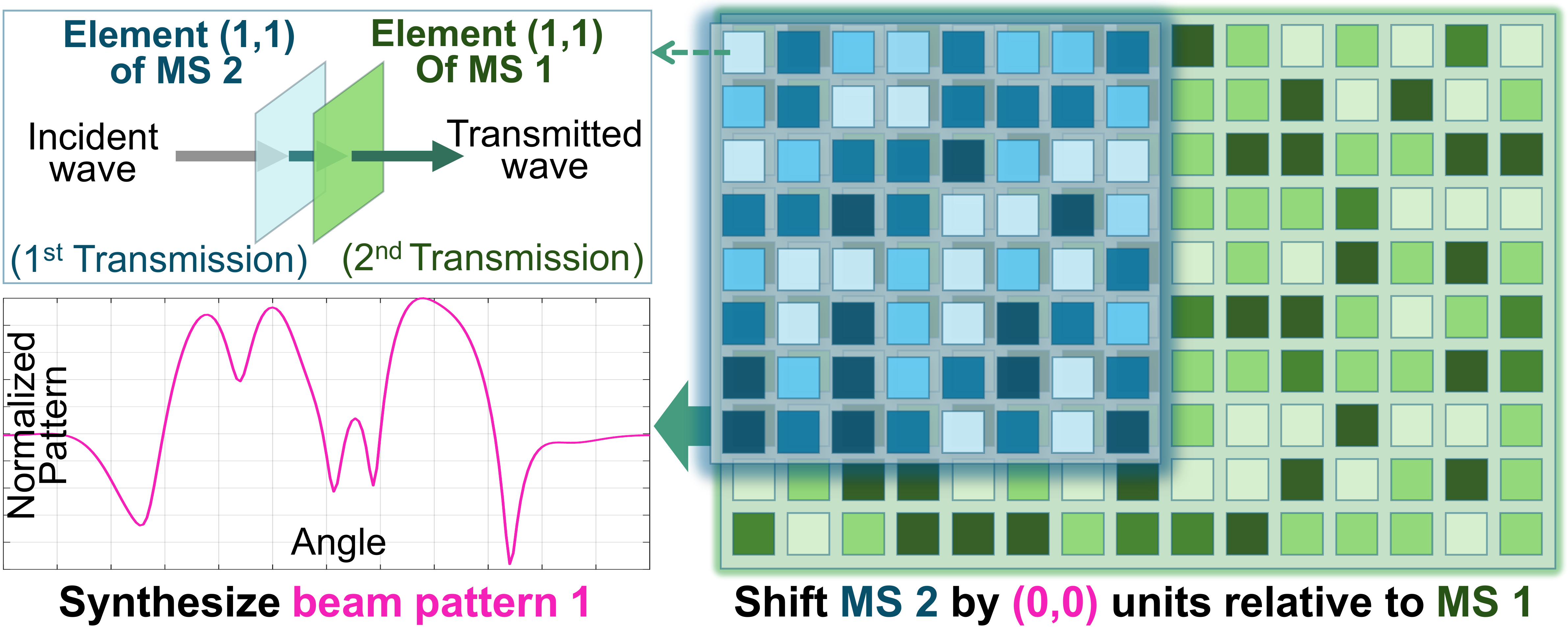}}
    \vspace{3pt}
    \subfloat[Shifting MS 2 by (3,1) units to synthesize beam pattern 2.]{
        \includegraphics[width=3.32in]{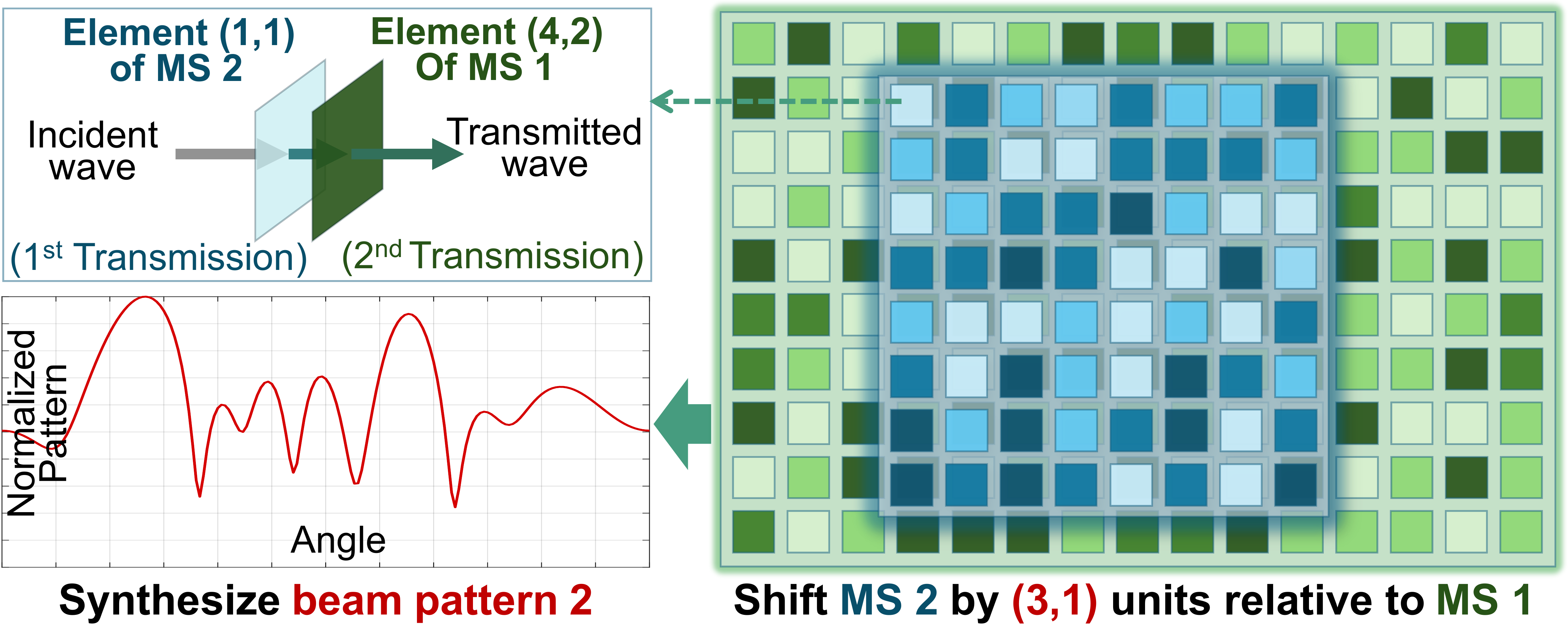}}
    \vspace{3pt}
    \subfloat[Shifting MS 2 by (6,2) units to synthesize beam pattern 3.]{
        \includegraphics[width=3.32in]{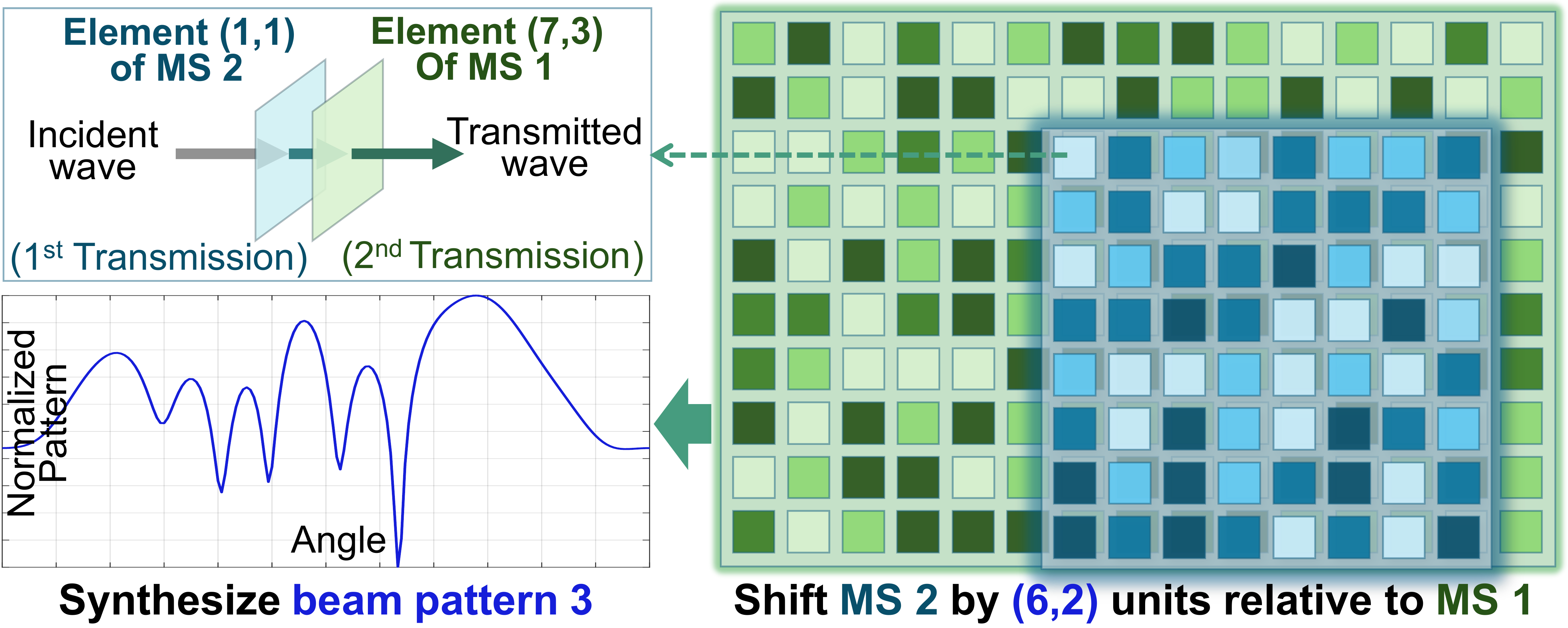}}
    \vspace{3pt}
    \captionsetup{font=small}
    \caption{An example of the proposed MIS architecture and its dynamic beam steering mechanism.}
    \vspace{-12pt}
\end{figure}

\subsection{MIS Beam Steering Mechanism}
To achieve dynamic beam steering, we introduce a method termed \textit{differential position-shifting}, which involves integrally shifting the position of MS 2 within MS 1 along its column or row directions by certain discrete units. By adjusting the position of MS 2, its transmissive elements align with different subsets of MS 1's elements. This realignment, combined with the superimposed phase shifts from successive transmissions, generates a new set of equivalent phase shifts, resulting in the synthesis of distinct beam patterns. Consequently, differential position shifting enables dynamic beam steering without the need for dynamic phase adjustments on MS 1 or MS 2. Fig. 2 shows an example of the proposed MIS architecture and its dynamic beam steering mechanism.
The beam steering process is summarized in the following steps:
\begin{enumerate}
    \item \textbf{Differential position shifting}: MS 2 shifts its position differentially within MS 1.
    \item \textbf{Transmissive elements re-aligning}: MS 2's elements align with different subsets of MS 1's elements.
    \item \textbf{Beam pattern synthesizing}: The superimposed phase shifts of MS 1 and MS 2 generate distinct beam patterns.
    \item \textbf{Dynamic Beam Steering}: By adjusting MS 2's positions, MIS switches beam patterns and steers beams to desired directions or locations within the target coverage area.
\end{enumerate}
This differential position shifting approach enables MIS beam steering with low hardware costs by eliminating the need for element-wise adjustments. Although MS 2 introduces an additional mobility requirement, it is adjusted as a whole within predefined discrete positions, avoiding the requirement for element-wise repositioning hardware or sophisticated position determination methods typical in MA systems [25]-[27]. 

Then, to quantitatively characterize the beamforming flexibility of the proposed MIS system, we define the total number of available beam patterns as $U$. This number is determined by the available discrete positions that MS 2 can occupy over MS 1, calculated based on the following parameters:
\begin{subequations}
\begin{align}
    U &= U_r \times U_c, \\
    U_r &= M_r - N_r + 1, \\
    U_c &= M_c - N_c + 1,
\end{align}
\end{subequations}
where $U_r$ and $U_c$ denote the number of discrete positions that MS 2 can shift along the row and column directions of MS 1, respectively. Each beam pattern $u \in \mathcal{U}$ corresponds to a unique overlap position of MS 2 on MS 1. Specifically, the beam pattern index $u$ captures MS 2's position shifts relative to MS 1 in both row and column directions, quantified by $u_r$ and $u_c$ units, respectively, such that
\begin{equation}
    u = (u_r - 1) \times U_c + u_c,
\end{equation}
where $u_r$ and $u_c$ denote the unit shifts in the row and column directions, respectively.

We emphasize that although this section outlines a specific MIS architecture, the MIS concept does not inherently require the contained MSs to vary in size or be restricted to rectangular shapes. For example, a MIS with uniformly sized MSs can also employ differential position shifting, and circular MSs can be relatively rotated. However, these movements may introduce modeling challenges, such as alterations in the effective aperture and difficulties in element alignment.

\section{MIS Communication Model and \\Problem Formulation}
\subsection{Channel Model}
Without loss of generality, we designate MS 1 as the first layer and MS 2 as the second layer of the MIS. This configuration implies that the signal transmitted by BS first arrives at MS 1, then sequentially
interacts with and passes through MS 1 and MS 2, and finally reaches the target coverage area.
Given that MS 1 and MS 2 are closely stacked, the signal propagation distance between them is negligible. Furthermore, we assume that both MS layers have a minimal thickness, ensuring that phase deviations introduced by the signal traversing the MIS are insignificant.

To clearly illustrate our proposed MIS architecture and beam steering mechanism, we adopt a basic channel model dominated by line-of-sight (LoS) propagation, which is typical in high-frequency bands such as millimeter waves. Consequently, the key channel matrices and vectors are defined as follows:
\begin{itemize}
    \item BS to MIS channel $\boldsymbol{G}$: Modeled as a deterministic LoS channel matrix $\boldsymbol{G} = \boldsymbol{a}_{\text{MIS}}\boldsymbol{a}_{\text{BS}}^T \in \mathbb{C}^{M \times L}$, where each element represents the complex channel coefficients from the $l$-th BS antenna to the $m$-th element of MS 1, derived from the array responses $\boldsymbol{a}_{\text{MIS}} \in \mathbb{C}^{M \times 1}$ and $\boldsymbol{a}_{\text{BS}} \in \mathbb{C}^{L \times 1}$. 
    \item MIS to $k$-th user channel $\boldsymbol{h}_k$: The channel vector $\boldsymbol{h}_k \in \mathbb{C}^{M \times 1}$ incorporates LoS components from the MIS elements, including those of MS 1 or MS 2, to $k$-th user. The specific elements included depend on the position of MS 2, comprising both the non-overlapping elements of the first layer MS 1 and the overlapping elements of the second layer MS 2.
\end{itemize}

The array response vectors involved in the above channel modeling are respectively defined as 
\begin{subequations}
\begin{align}
&\!\! \boldsymbol{a}_{\text{MIS}}\left( \vartheta _{\text{MIS}},\psi _{\text{MIS}} \right) \nonumber
\\
&\!\!=[1,e^{j \frac{2\pi d_{\text{MIS}}}{\lambda}\left( m_r\cos \left( \vartheta _{\text{MIS}} \right) \sin \left( \psi _{\text{MIS}} \right) +m_c\sin \left( \vartheta _{\text{MIS}} \right) \sin \left( \psi _{\text{MIS}} \right) \right)},\dots, \nonumber
\\
&\!\!\,\, e^{j \frac{2\pi d_{\text{MIS}}}{\lambda}\left( \left( M_r\!-\!1 \right) \cos \left( \vartheta _{\text{MIS}} \right) \sin \left( \psi_{\text{MIS}} \right) +\left( M_c\!-\!1 \right) \sin \left( \vartheta_{\text{MIS}} \right) \sin \left( \psi_{\text{MIS}} \right) \right)}]^T\!,\!\!
\end{align}
\vspace{-16pt}\begin{align}
&\!\!\boldsymbol{a}_{\text{BS}}\left( \vartheta _{\text{BS}},\psi _{\text{BS}} \right)
\nonumber
\\
&\!\!=[1,e^{j \frac{2\pi d_{\text{BS}}}{\lambda}\left( n_r\cos \left( \vartheta _{\text{BS}} \right) \sin \left( \psi _{\text{BS}} \right) +n_c\sin \left( \vartheta _{\text{BS}} \right) \sin \left( \psi _{\text{BS}} \right) \right)},\dots , \nonumber
\\
&\!\!\,\,e^{j2\pi \frac{2\pi d_{\text{BS}}}{\lambda}\left( \left( L_r\!-\!1 \right) \cos \left( \vartheta _{\text{BS}} \right) \sin \left( \psi _{\text{BS}} \right) +\left( L_c\!-\!1 \right) \sin \left( \vartheta _{\text{BS}} \right) \sin \left( \psi _{\text{BS}} \right) \right)}]^T\!,\!\! 
\\
&\!\!\boldsymbol{h}_k\left( \vartheta _k,\psi _k \right) \nonumber
\\
&\!\!=[1,e^{j2\pi \frac{2\pi d_{\text{MIS}}}{\lambda}\left( n_r\cos \left( \vartheta _k \right) \sin \left( \psi _k \right) +n_c\sin \left( \vartheta _k \right) \sin \left( \psi _k \right) \right)},\dots , \nonumber
\\
&\!\!\,\,e^{j2\pi \frac{2\pi d_{\text{MIS}}}{\lambda}\left( \left( L_r-1 \right) \cos \left( \vartheta _k \right) \sin \left( \psi _k \right) +\left( L_c-1 \right) \sin \left( \vartheta _k \right) \sin \left( \psi _k \right) \right)}]^T,
\end{align}    
\end{subequations}
where $d_{\text{BS}}$ denotes the antenna spacing for the BS, configured as a uniform planar array (UPA). The angles $(\vartheta_{\text{BS}}, \psi_{\text{BS}})$, $(\vartheta_{\text{MIS}}, \psi_{\text{MIS}})$, and $(\vartheta_k, \psi_k)$ represent the azimuth and elevation angles of departure (AoD) for the BS, the azimuth and elevation angles of arrival (AoA) for the MIS, and the azimuth and elevation AoA for the direction of $k$-th user, respectively.

\vspace{-10pt}
\subsection{Transmission Signal Model}
\vspace{-2pt}
Building on the channel model, the proposed MIS architecture, and its beam steering mechanism, we present the MIS-enabled transmission signal model. Signals transmitted by the BS can be coordinately refracted by both MS 1 and MS 2 through their overlapping elements or solely by MS 1 through its non-overlapping elements. Since MS 1 is fixed, its signal representation mirrors that of conventional RIS-aided wireless systems. The primary challenge lies in accurately modeling the movable MS 2. 
To explicitly capture the impact of position shifting on the $u$-th beam pattern, we define a set of $U$ equivalent phase shift vectors for MS 2 as follows:
\begin{equation}
\boldsymbol{\bar{\theta}}_{u}= \boldsymbol{S}_u \boldsymbol{\theta} + \boldsymbol{e}_u \in \mathbb{C}^{M\times 1}, \forall u \in \mathcal{U}, 
\end{equation}
where we introduce a set of binary selection matrices $\{\boldsymbol{S}_u\in \{0,1\}^{M \times N}, \forall u\in\mathcal{U}\}$ and a set of padding vectors $\{\boldsymbol{e}_u\in \{0,1\}^{M \times 1},\forall u\in\mathcal{U}\}$, defined as below:
\begin{subequations}
\begin{align}
\left[ \boldsymbol{S}_u \right] _{m,n}&=\begin{cases}
	1, &	\!\!\!\begin{array}{c}
	\text{if }n\text{-th element of MS 2 locates}\\
	\!\!\!\!\!\!\!\text{upon }m\text{-th element of MS 1},\\
\end{array}\\
	0,&		\text{otherwise},\\
\end{cases}
\\
\left[ \boldsymbol{e}_u \right]_m&=\begin{cases}
	0, &	\!\!\!\begin{array}{c}
	\text{if }n\text{-th element of MS 2 locates}\\
	\!\!\!\!\!\!\!\text{upon }m\text{-th element of MS 1},\\
\end{array}\\
	1,&		\text{otherwise}.\\
\end{cases}
\end{align}
\end{subequations}

Specifically, the binary selection matrix $\boldsymbol{S}_u$ identifies the discrete positions on MS 1 where the elements of MS 2 are superimposed, defining the overlapping relationship between the position-shifted MS 2 and the subsets of MS 1. Additionally, the binary padding vector $\boldsymbol{e}_u$ contains $M-N$ entries set to one, representing virtual MS 2 elements with zero phase that are added to the non-overlapping regions of MS 1. Note that this extension aligns the equivalent size of MS 2 with the size of MS 1, facilitating simplified signal representation. 

The BS employs beamforming vectors to transmit symbols to users with specific directions or locations. For the $k$-th user, the transmit signal is given by $\boldsymbol{x_k} = \boldsymbol{w}_k s_k$, where $\boldsymbol{w}_k \in \mathbb{C}^{M \times 1}$ is the beamforming vector and $s_k \in \mathbb{C}$ is the transmitted symbol with $\mathbb{E}[|s_k|^2] = 1$. The transmit power is constrained by $\|\boldsymbol{w}_k\|^2 \leq P_{\text{max}}, \quad \forall k \in \mathcal{K}$ with maximum transmit power $P_{\text{max}}$. Accordingly, the received signal at $k$-user under $u$-th beam pattern is modeled as
\begin{subequations}
\begin{align}
y_k&=\boldsymbol{h}_{k}^{T}\mathrm{diag}(\boldsymbol{\bar{\theta}}_u) \mathrm{diag}\left( \boldsymbol{\phi } \right) \boldsymbol{Gw}_{k,u}s_u +n_k
\\
&=\boldsymbol{h}_{k}^{T}\boldsymbol{\bar{\varTheta}}_{u}\boldsymbol{\varPhi G}\boldsymbol{w}_{k,u}s_u +n_k,
\end{align}
\end{subequations}
where $\boldsymbol{\bar{\varTheta}}_{u}=\mathrm{diag}( \boldsymbol{\bar{\theta}}_u )$ represents the equivalent diagonal transmissive matrix of MS 2 with $u$-th shifted position, and $n_k \sim \mathcal{CN}(0, \sigma^2)$ denotes the additive white gaussian noise (AWGN) at user $k$. Note that the model of the interaction between MSs in MIS, i.e., (4) and (5), as well as the MIS transmission signal model (6), can be generally extended to multi-layer MIS architectures.

For beam coverage design in our considered MIS-aided system, which is similar to multiple-input single-output (MISO) communications, the BS employs maximum-ratio transmission (MRT) beamforming $\boldsymbol{w}_{k,u}=\sqrt{P_{\max}}\frac{( \boldsymbol{h}_{k}^{T}\boldsymbol{\bar{\varTheta}}_u\boldsymbol{\varPhi G}) ^H}{\lVert \boldsymbol{h}_{k}^{T}\boldsymbol{\bar{\varTheta}}_u\boldsymbol{\varPhi G} \rVert}$ to achieve optimal beamforming gain for target directions of users within the target area.
Under this context, the received SNR at $k$-th user served by $u$-th beam pattern can be expressed as
\begin{subequations}
\begin{align}
\gamma _{k,u}\left( \boldsymbol{\phi },\boldsymbol{\bar{\theta}}_u \right)&=\frac{P_{\max}}{\sigma _{k}^{2}}\left\| \boldsymbol{h}_{k}^{T}\boldsymbol{\bar{\varTheta}}_{u}\boldsymbol{\varPhi G} \right\|^2\frac{1}{\sigma _{k}^{2}}
\\
&=\frac{P_{\max}L}{\sigma _{k}^{2}}\left| \left( \boldsymbol{\bar{\theta}}_{u}\odot \boldsymbol{\phi } \right) ^T\mathrm{diag}\left( \boldsymbol{h}_{k}^{} \right) \boldsymbol{a}_{\text{MIS}} \right|^2
\\
&=\iota_k| \left( \boldsymbol{\bar{\theta}}_u\odot \boldsymbol{\phi } \right) ^T\boldsymbol{c}_k |^2,
\end{align}
\end{subequations}
where $\boldsymbol{c}_k=\mathrm{diag}\left( \boldsymbol{h}_{k}^{} \right) \boldsymbol{a}_{\text{MIS}}$ denotes the two-hop cascaded channel from the BS to $k$-user enabled by the MIS, and we denote $\iota_k=\frac{P_{\max}L}{\sigma _{k}^{2}}$ for brevity.

\subsection{Joint MIS Phase Shifts Design and Position Selection Problem Formulation}
\begin{figure}[t]
    \centering
    \includegraphics[width=2.6in]{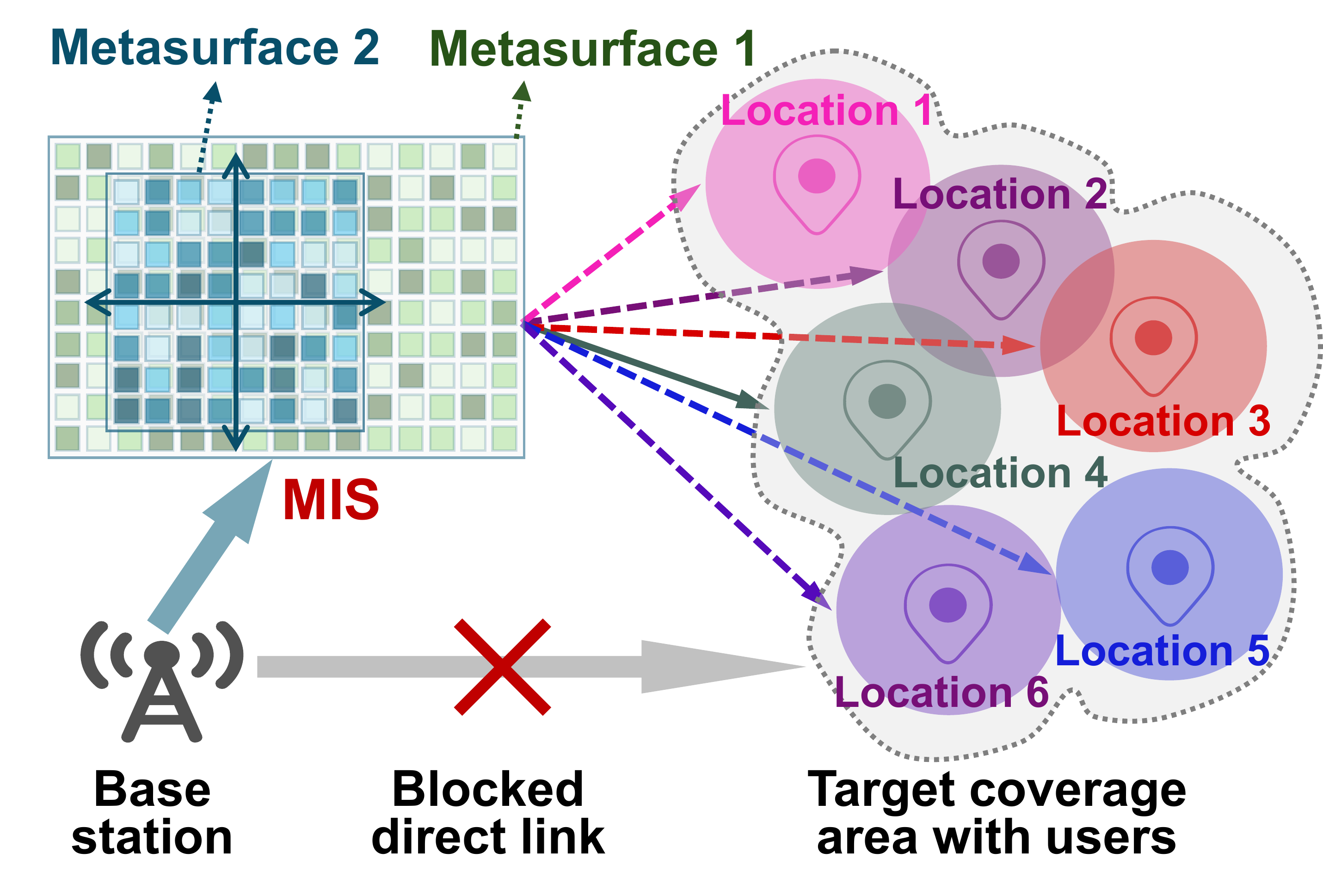}
    \captionsetup{font=small}
    \caption{Illustration of the MIS-aided coverage extension scenario.} 
    \vspace{-12pt}
    \label{fig:system_model}
\end{figure}
As shown in Fig. 3, in the scenario of MIS-aided coverage extension, the goal of system design is to maximize the worst-case received SNR across all designated directions (or equivalently, the minimum received SNR among all users) within the target area $\mathcal{A}$. This is achieved by jointly optimizing the static phase shifts of the MIS, including $\boldsymbol{\varPhi}$ for MS 1 and $\boldsymbol{\varTheta}$ for MS 2, as well as selecting the optimal position $u^{\text{opt}}$ for MS 2 when serving $k$-th user. Accordingly, the joint optimization problem is formulated as follows:
\begin{subequations}
\begin{align}
    \text{(P1):}\,\,\max_{\boldsymbol{\phi}, \boldsymbol{\theta}, \boldsymbol{X}} \,\, & \min_{k \in \mathcal{K}} \,\, \bigg\{ \sum_{u=1}^{U} x_{k,u} \gamma_{k,u}\left( \boldsymbol{\phi}, \boldsymbol{\bar{\theta}}_u \right) \bigg\} \\
    \text{s.t.} \,\,& \sum_{u=1}^{U} x_{k,u} = 1, \, \forall k \in \mathcal{K}, \\
    & x_{k,u} \in \{0,1\}, \, \forall k \in \mathcal{K}, \forall u \in \mathcal{U}, \\
    & |\phi_m| = 1, \, \forall m \in \mathcal{M}, \\
    & |\theta_n| = 1, \, \forall n \in \mathcal{N}.
\end{align}
\end{subequations}
where $\boldsymbol{X} \in \{0,1\}^{K \times U}$ is the binary beam pattern scheduling matrix with $x_{k,u} = 1$ indicating that $k$-th user is assigned to $u$-th beam pattern. Note that the determination of the index $u$ corresponds to the selection of the position to shift for MS 2 as defined in (2). Constraints (8b) and (8c) ensure that each user is assigned exactly one beam pattern from the available set $\mathcal{U}$, while allowing a beam pattern to be selected by multiple users as needed. In summary, the beam pattern scheduling matrix $\boldsymbol{X}$ represents the position selection of MS 2, enabling each user to be adaptively assigned the most suitable beam pattern.

Now, we focus on developing an efficient algorithm to solve the formulated problem (P1). This optimization problem is inherently non-convex due to the unit-modulus constraints on the MIS phase shifts, the binary nature of the beam pattern scheduling variables, and the max-min objective function with coupled optimization variables.\footnote{We adopt a generalized formulation to emphasize the key features and degrees of freedom of the MIS, while recognizing that its operation can be simplified in practice. For instance, restricting MS 2's movement to be 1D can reduce movement hardware costs. Alternatively, we may simplify position selection by directly mapping each beam pattern to a user in a predefined one-to-one assignment. These simplified approaches offer a balanced trade-off between the complexity and flexibility of the MIS architecture.} 
Consequently, it is challenging to solve directly using conventional methods.
To effectively tackle this problem, we employ the Riemannian manifold optimization framework, treating the unit-modulus constraints as a manifold feasible region. However, manifold optimization algorithms typically require a smooth objective function and manifold constraints. Therefore, additional challenges include handling the non-smooth max-min objective function and the discrete 0-1 beam pattern scheduling variables. In the following section, we address these obstacles to derive a solution for (P1).

\section{Proposed Algorithm for Problem (P1)}
In this section, we develop a tailored manifold optimization algorithm that jointly designs the phase shifts $\boldsymbol{\phi}$ and $\boldsymbol{\theta}$ of MS 1 and MS 2, respectively, along with the beam pattern scheduling matrix $\boldsymbol{X}$, aiming to maximize the worst-case received SNR among all users.
Given the mixed-integer non-smooth nature of (P1), we first replace the non-smooth max-min function with a smooth approximation and then relax the binary scheduling variables to construct a multinomial manifold. These modifications result in a smooth objective function with manifold constraints, which makes the reformulated problem suitable for manifold optimization methods.

\subsection{Smoothing for the non-smooth Max-Min Function}
The manifold optimization framework requires that the objective function be smooth. However, the max-min form in the objective function of (P1) is inherently non-smooth, posing challenges for gradient-based methods. To address this, we employ the log-sum-exponential (LSE) smoothing technique. 

Specifically, by introducing a smoothing parameter $\mu > 0$, we transform the original objective function into a smooth approximation $f\left( \boldsymbol{\phi}, \boldsymbol{\theta}, \boldsymbol{X} \right)$ with infinite differentiability
\begin{align}
f\left( \boldsymbol{\phi },\boldsymbol{\theta},\boldsymbol{X} \right) =-\mu \log \left( \sum_{k=1}^K{\exp}\left( -\frac{g_k\left(\boldsymbol{\phi},\boldsymbol{\theta},\boldsymbol{X}\right)}{\mu} \right) \right) ,
\end{align}
where we denote the set of objective user SNR values to be minimized over $\mathcal{K}$ as 
\begin{align}
g_k\left(\boldsymbol{\phi},\boldsymbol{\theta},\boldsymbol{X}\right) =\sum_{u=1}^U{x_{k,u}}\gamma _{k,u}\left( \boldsymbol{\phi },\boldsymbol{\bar{\theta}}_u \right).
\end{align}
The smoothed function $f\left( \boldsymbol{\phi}, \boldsymbol{\theta}, \boldsymbol{X} \right)$ approximates the original max-min function $\min_{k\in \mathcal{K}} \{ g_k\left(\boldsymbol{\phi}, \boldsymbol{\theta}, \boldsymbol{X}\right) \}$ in (8a) and satisfies
\begin{align}
&\!\!\!\! f\left( \boldsymbol{\phi },\boldsymbol{\theta},\boldsymbol{X} \right) \le \min_{k\in \mathcal{K}} \{ g_k\left(\boldsymbol{\phi},\boldsymbol{\theta},\boldsymbol{X}\right)\}  \le f\left( \boldsymbol{\phi },\boldsymbol{\theta},\boldsymbol{X} \right)\!+\!\mu \log K, \!\!
\end{align}
which indicates that the smoothing parameter $\mu$ controls the accuracy of the approximation [28]. As $\mu$ approaches zero or becomes sufficiently small, the inequality tightens, and the surrogate function closely approximates the minimum SNR.

However, an excessively small $\mu$ may lead to an ill-conditioned optimization problem with numerical instability. To maintain stability while preserving high accuracy, we adopt an iterative strategy that progressively reduces $\mu$, solving a sequence of problems with increasingly accurate approximations. Specifically, starting with an initial value $\mu^{(0)}$, we iteratively update $\mu$ in $t$-th iteration as
\begin{align}
\mu^{(t)} = \mu^{(t-1)}\delta ^{-1},
\end{align}
where the update-rate parameter $\delta > 1$ is typically chosen as $\delta = 2$ \cite{ref28}. This progressively refining approach allows the LSE approximation to closely resemble the original max-min function. Thus, we convert the original non-smooth objective function of (P1) into a smooth one that is amenable to Riemannian gradient-based manifold optimization techniques.

\subsection{Product Manifold Constructing}
\label{subsec:Manifold_Feasible_Region}
After smoothing the objective function, we focus on constructing a manifold for the optimization variables. The phase shifts $\boldsymbol{\phi}$ and $\boldsymbol{\theta}$ are subject to unit-modulus constraints, i.e., $|\phi_m| = 1, \forall m \in \mathcal{M}$ and $|\theta_n| = 1, \forall n \in \mathcal{N}$, inherently defining a Riemannian manifold known as the \textit{complex circle manifold}.

The main challenge lies with the beam pattern scheduling variables $x_{k,u}$, which are binary and do not belong to any manifolds. To tackle this difficulty, we relax these variables to continuous values
\begin{align}
x_{k,u}\in \left\{ 0,1 \right\} \Rightarrow 1\ge x_{k,u}\ge 0, \,  \forall k \in \mathcal{K}, \forall u \in \mathcal{U}.
\end{align}
For the relaxed continuous variable $\boldsymbol{x}$ in (13), combined with the original equality multinomial constraints $\sum_{u=1}^U{x_{k,u}}=1$ in (8b), its range of values is naturally confined to be less than or equal to 1. Furthermore, we enforce strict positivity to ensure that:
\begin{align}
\sum_{u=1}^U{x_{k,u}}=1, x_{k,u} > 0, \,  \forall k \in \mathcal{K}, \forall u \in \mathcal{U},
\end{align}
which classifies the relaxed beam pattern scheduling variables into a multinomial manifold [29].
The optimal value of the relaxed beam pattern scheduling variables $x_{k,u}$ in (P1) inherently converges to binary values with only a trivial gap due to the strict inequality $x_{k,u} > 0$. This convergence occurs because, for each user $k$, the optimization favors $x_{k,u}$ being one for the beam patterns $u$ with the highest $\gamma_{k,u}$ and being zero otherwise, effectively driving $x_{k,u}$ towards binary values in the limit $x_{k,u} \in \{0,1\}$.

Now, we summarize the manifolds of the optimization variables, i.e., the phase shift vectors $\boldsymbol{\phi}$ and $\boldsymbol{\theta}$ lie on the complex circle manifold and the beam pattern scheduling matrix $\boldsymbol{X}$ lies on the multinomial manifold, as follows
\begin{subequations}
\begin{align}
    &\mathcal{R}_{\boldsymbol{\phi}} = \left\{ \boldsymbol{\phi} \in \mathbb{C}^M : |\phi_m| = 1, \forall m \in \mathcal{M} \right\}, 
    \\
    &\mathcal{R}_{\boldsymbol{\theta}} = \left\{ \boldsymbol{\theta} \in \mathbb{C}^N : |\theta_n| = 1, \forall n \in \mathcal{N}\right\}, 
    \\
    &\mathcal{R}_{\boldsymbol{X}} \! =\! \Big\{ \!\boldsymbol{X} \in \mathbb{R}^{K \times U}\!:\! \sum_{u\in \mathcal{U}} x_{k,u}\! =\!1, x_{k,u} \!>\!0, \forall u \in \mathcal{U}, \forall k \in \mathcal{K}  \Big\}.
\end{align}
\end{subequations}
For intuitive understanding, a manifold can be interpreted as a topological space that locally resembles the Euclidean space. Then, a tangent vector describes the direction in which a point can be updated on the manifold. All tangent vectors at a given point, representing all possible directions in which the point can move, collectively form the tangent space [16]. Therefore, each tangent space can be regarded as a Euclidean space and is equipped with tangent vectors, including the Riemannian gradient, which points in the direction where the objective function decreases most rapidly [29]. The corresponding tangent space \(\mathcal{T}\) for \(\boldsymbol{\phi}\) and \(\boldsymbol{\theta}\) at their respective local points can be respectively given by:
\begin{subequations}
\begin{align}
&\mathcal{T}_{\boldsymbol{\phi }}=\big\{\boldsymbol{t}\in \mathbb{C}^M: \Re\left\{ \phi _m^* t_m \right\} =0,\forall m\in \mathcal{M}\big\},
\\
&\mathcal{T}_{\boldsymbol{\theta }}=\big\{\boldsymbol{t}\in \mathbb{C}^N: \Re\left\{ \theta _m^* t_m \right\} =0,\forall n\in \mathcal{N}\big\}.
\end{align}
For $\boldsymbol{X}$ on the multinomial manifold, the tangent space consists of matrices where the elements in each row sum to zero, defined as
\begin{align}
\mathcal{T}_{\boldsymbol{X}}=\Big\{ \boldsymbol{T}\in \mathbb{R}^{K\times U}: \sum_{u \in \mathcal{U}}{t_{k,u}}=0,\forall u\in \mathcal{U}, \forall k\in \mathcal{K} \Big\},\!\!
\end{align}
\end{subequations}
where condition $\sum_{u \in \mathcal{U}}{t_{k,u}}=0$ ensures that any infinitesimal perturbation within the tangent space does not violate the simplex constraint that the components sum to one.

Subsequently, with these definitions, we jointly optimize these variables while complying with their respective manifold constraints. By taking the Cartesian product of the individual manifolds for optimization variables, we construct a product manifold defined as 
\begin{align}
    \mathcal{R}_{(\boldsymbol{\phi},\boldsymbol{\theta},\boldsymbol{X})} = \mathcal{R}_{\boldsymbol{\phi}} \times \mathcal{R}_{\boldsymbol{\theta}} \times \mathcal{R}_{\boldsymbol{X}},
\end{align}
where \(\times\) denotes the Cartesian product between sets. For any point \((\boldsymbol{\phi}, \boldsymbol{\theta}, \mathbf{X}) \in \mathcal{R}_{(\boldsymbol{\phi},\boldsymbol{\theta},\boldsymbol{X})}\), the tangent space at that point is the direct sum of the individual tangent spaces
\begin{align}
    \mathcal{T}_{(\boldsymbol{\phi},\boldsymbol{\theta},\mathbf{X})} = \mathcal{T}_{\boldsymbol{\phi}} \oplus \mathcal{T}_{\boldsymbol{\theta}} \oplus \mathcal{T}_{\mathbf{X}},
\end{align}
where \(\oplus\) denotes the direct sum of vector spaces.

\subsection{Riemannian Conjugate Gradient Method}
By recasting (P1) over the defined product manifold with the smoothed objective function, we formulate it as a standard Riemannian manifold optimization problem 
\begin{align}
\text{(P2):}\,\,\max_{\boldsymbol{\phi},\boldsymbol{\theta},\boldsymbol{X}\in\mathcal{R}_{(\boldsymbol{\phi},\boldsymbol{\theta},\boldsymbol{X})}}&\,\,f\left( \boldsymbol{\phi },\boldsymbol{\theta},\boldsymbol{X} \right)
\end{align}
This formulation allows us to jointly optimize the three variables within a unified manifold optimization framework without resorting to the relaxation of unit-modulus constraints or employing block coordinate descent (BCD) procedures, thereby achieving favorable convergence performance while taking the constraints and interactions of all variables into consideration.

To solve the reformulated manifold optimization problem, we employ the Riemannian Conjugate Gradient (RCG) method. The RCG method is an extension of the classical conjugate gradient algorithm to Riemannian manifolds, allowing efficient optimization over curved spaces. The key steps in the RCG method involve determining the search direction to update the variables and performing a retraction to ensure that the variables remain in the manifold. In the following section, we provide a detailed description of the algorithm's construction steps.

\subsubsection{Computing the Riemannian gradient}
Before determining the search direction, we need to calculate the Riemannian gradient, which is a vector field on the manifold \( \mathcal{R} \) obtained by projecting the Euclidean gradient onto the tangent space of the manifold. Specifically, using the chain rule on the multivariate composite function with functional relationships referred to (4), (7), (9), and (10), the Euclidean gradients of the objective function $f(\boldsymbol{\phi}, \boldsymbol{\theta}, \boldsymbol{X})$ with respect to the optimization variables $\boldsymbol{\phi}$, $\boldsymbol{\theta}$ and $\boldsymbol{X}$ are, respectively, given by
\begin{subequations}
\begin{align}
&\!\!\nabla _{\boldsymbol{\phi }}f\left( \boldsymbol{\phi },\boldsymbol{\theta },\boldsymbol{X} \right) \nonumber
\\
&\!\!=\sum_{k=1}^K{v_k}\left( \boldsymbol{\phi },\boldsymbol{\theta },\boldsymbol{X} \right) \nabla _{\boldsymbol{\phi }}g_k\left( \boldsymbol{\phi },\boldsymbol{\theta },\boldsymbol{X} \right) \nonumber
\\
&\!\!=\sum_{k=1}^K{\sum_{u=1}^U{v_k\left( \boldsymbol{\phi },\boldsymbol{\theta },\boldsymbol{X} \right) x_{k,u}\nabla _{\boldsymbol{\phi }}\gamma _{k,u}\left( \boldsymbol{\phi },\boldsymbol{\bar{\theta}}_u \right)}} \nonumber
\\
&\!\!=\sum_{k=1}^K{\sum_{u=1}^U{2\iota _kv_k\left( \boldsymbol{\phi },\boldsymbol{\theta },\boldsymbol{X} \right) x_{k,u}q_{k,u}\left( \boldsymbol{\phi },\boldsymbol{\bar{\theta}}_u \right) \big( \mathrm{diag}\left( \boldsymbol{\bar{\theta}}_u \right) \boldsymbol{c}_k \big) ^*}},
\\
&\!\!\nabla _{\boldsymbol{\theta }}f\left( \boldsymbol{\phi },\boldsymbol{\theta },\boldsymbol{X} \right)  \nonumber
\\
&\!\!=\!\sum_{k=1}^K{\!}v_k\left( \boldsymbol{\phi },\!\boldsymbol{\theta },\!\boldsymbol{X} \right) \!\sum_{u=1}^U{\!}x_{k,u}\nabla _{\boldsymbol{\theta }}\gamma _{k,u}\left( \boldsymbol{\phi },\!\boldsymbol{\bar{\theta}}_u \right) \nonumber
\\
&\!\!=\!\sum_{k=1}^K{\sum_{u=1}^U{2}}\iota_k v_k\!\left( \boldsymbol{\phi },\boldsymbol{\theta },\boldsymbol{X} \right) x_{k,u}q_{k,u}\!\left( \boldsymbol{\phi },\boldsymbol{\bar{\theta}}_u \right)\!\! \big( \boldsymbol{S}_u^T\mathrm{diag}( \boldsymbol{\phi }) \boldsymbol{c}_k \big)^*,
\\
&\!\!\nabla _{\boldsymbol{X}}f\left( \boldsymbol{\phi },\boldsymbol{\theta },\boldsymbol{X} \right) =\left[ \frac{\partial f\left( \boldsymbol{\phi },\boldsymbol{\theta },\boldsymbol{X} \right)}{\partial x_{k,u}} \right] _{K\times U} \nonumber
\\
&\quad \quad \quad \quad \quad \quad  =\left[ v_k\left( \boldsymbol{\phi },\boldsymbol{\theta },\boldsymbol{X} \right) \gamma _{k,u}\left( \boldsymbol{\phi },\boldsymbol{\bar{\theta}}_u \right) \right] _{K\times U},
\end{align}
\end{subequations}
where we define
\begin{subequations}
\begin{align}
&v_k(\boldsymbol{\phi}, \boldsymbol{\theta}, \boldsymbol{X}) = \dfrac{\exp\left( -\dfrac{g_k(\boldsymbol{\phi}, \boldsymbol{\theta}, \boldsymbol{X})}{\mu} \right)}{\sum_{i=1}^K \exp\left( -\dfrac{g_i(\boldsymbol{\phi}, \boldsymbol{\theta}, \boldsymbol{X})}{\mu} \right)},
\\
&q_{k,u}\left( \boldsymbol{\phi },\boldsymbol{\bar{\theta}}_u \right) = \left( \boldsymbol{\bar{\theta}}_u \odot \boldsymbol{\phi} \right)^\mathrm{T} \boldsymbol{c}_k.
\end{align}
\end{subequations}
Having computed the Euclidean gradients of the objective function $f(\boldsymbol{\phi}, \boldsymbol{\theta}, \boldsymbol{X})$ with respect to the optimization variables, we project them onto the tangent spaces of their respective manifolds to obtain the Riemannian gradients.

\textit{Riemannian Gradient with respect to \(\boldsymbol{\phi}\):} The Riemannian gradient \(\nabla_{\mathcal{R}_{\boldsymbol{\phi}}} f(\boldsymbol{\phi}, \boldsymbol{\theta}, \boldsymbol{X})\) is obtained by projecting the Euclidean gradient \(\nabla_{\boldsymbol{\phi}} f(\boldsymbol{\phi}, \boldsymbol{\theta}, \boldsymbol{X})\) onto the tangent space \(\mathcal{T}_{\boldsymbol{\phi}}\)
\begin{subequations}
\begin{align}
&\nabla_{\mathcal{R}_{\boldsymbol{\phi}} } f(\boldsymbol{\phi}, \boldsymbol{\theta}, \boldsymbol{X}) \nonumber
\\
&= \mathsf{Proj}_{\boldsymbol{\phi}}(\nabla_{\boldsymbol{\phi}} f(\boldsymbol{\phi}, \boldsymbol{\theta}, \boldsymbol{X})) 
\\
&= \nabla_{\boldsymbol{\phi}} f(\boldsymbol{\phi}, \boldsymbol{\theta}, \boldsymbol{X}) - \Re\left( \nabla_{\boldsymbol{\phi}} f(\boldsymbol{\phi}, \boldsymbol{\theta}, \boldsymbol{X}) \odot \boldsymbol{\phi}^* \right) \odot \boldsymbol{\phi},
\end{align}
\end{subequations}
where $\mathsf{Proj}_{\boldsymbol{\phi}}(\cdot)$ denotes the projection operation, $\odot$ denotes the Hadamard (element-wise) product, and $\boldsymbol{\phi}^*$ is the complex conjugate of $\boldsymbol{\phi}$.

\textit{Riemannian Gradient with respect to \(\boldsymbol{\theta}\):} Similarly, the Riemannian gradient \(\nabla_{\mathcal{R}_{\boldsymbol{\theta}}} f(\boldsymbol{\phi}, \boldsymbol{\theta}, \boldsymbol{X})\) is given by
\begin{subequations}
\begin{align}
\nabla_{\mathcal{R}_{\boldsymbol{\theta}} }& f(\boldsymbol{\phi}, \boldsymbol{\theta}, \boldsymbol{X}) \nonumber
\\
=& \mathsf{Proj}_{\boldsymbol{\theta}}(\nabla_{\boldsymbol{\theta}} f(\boldsymbol{\phi}, \boldsymbol{\theta}, \boldsymbol{X})) 
\\
=& \nabla_{\boldsymbol{\theta}} f(\boldsymbol{\phi}, \boldsymbol{\theta}, \boldsymbol{X}) - \Re\left( \nabla_{\boldsymbol{\theta}} f(\boldsymbol{\phi}, \boldsymbol{\theta}, \boldsymbol{X}) \odot \boldsymbol{\theta}^* \right) \odot \boldsymbol{\theta}.
\end{align}
\end{subequations}

\textit{Riemannian Gradient with respect to \(\boldsymbol{X}\):} For the beam pattern scheduling matrix \(\boldsymbol{X} \in \mathbb{R}^{K \times U}\), which lies on the multinomial manifold, the Riemannian gradient \(\nabla_{\mathcal{R}_{\boldsymbol{X}}} f(\boldsymbol{\phi}, \boldsymbol{\theta}, \boldsymbol{X})\) is obtained by 
\begin{subequations}
\begin{align}
&\nabla_{\mathcal{R}_{\boldsymbol{X}} } f(\boldsymbol{\phi}, \boldsymbol{\theta}, \boldsymbol{X}) \nonumber
\\
&= \mathsf{Proj}_{\boldsymbol{X}}(\nabla_{\boldsymbol{X}} f(\boldsymbol{\phi}, \boldsymbol{\theta}, \boldsymbol{X})) 
\\
&= \nabla_{\boldsymbol{X}} f(\boldsymbol{\phi}, \boldsymbol{\theta}, \boldsymbol{X}) - \Big( \frac{1}{U}\nabla_{\boldsymbol{X}} f(\boldsymbol{\phi}, \boldsymbol{\theta}, \boldsymbol{X}) \boldsymbol{1}_{U\times1} \Big) \boldsymbol{1}_{U\times1}^\mathrm{T},\!\!
\end{align}
\end{subequations}
where $\boldsymbol{1}_{U\times1}$ is a column vector of length $U$, and the subtraction ensures that each row of $\nabla_{\mathcal{R}_{\boldsymbol{X}} } f(\boldsymbol{\phi}, \boldsymbol{\theta}, \boldsymbol{X})$ sums to zero, satisfying the tangent space conditions in (16c) of the multinomial manifold in a probability simplex form.
These Riemannian gradients (22), (23), and (24) are then utilized in the manifold optimization algorithm to iteratively update the variables \(\boldsymbol{\phi}\), \(\boldsymbol{\theta}\), and \(\boldsymbol{X}\), respectively.

\subsubsection{Determining Conjugate Descent Direction}
To determine the descent direction on the manifold, we employ the conjugate gradient method adapted to Riemannian manifolds. Specifically, in iteration $i$, the descent direction $ \boldsymbol{\eta }^{\left( i \right)}$, $\boldsymbol{\tau }^{\left( i \right)}$, and $\boldsymbol{\varXi }^{\left( i \right)}$ for the variables $\boldsymbol{\phi}^{(i)}$, $\boldsymbol{\theta}^{(i)}$, and $\boldsymbol{X}^{(i)}$, respectively, are calculated using the Polak-Ribiere formula.

\textit{Descent Direction for \(\boldsymbol{\phi}\):} 
Let $\nabla _{\mathcal{R}_{\boldsymbol{\phi }}}^{}f(\boldsymbol{\phi}^{(i)},\boldsymbol{\theta}^{(i)},\boldsymbol{X}^{(i)})$ denote the Riemannian gradient in iteration $i$. For brevity, the function $f(\boldsymbol{\phi}^{(i)},\boldsymbol{\theta}^{(i)},\boldsymbol{X}^{(i)})$ is abbreviated as $f^{(i)}$ in the following. The conjugate gradient update coefficient $\beta_{\boldsymbol{\phi}}^{(i)}$, which adjusts the new search direction by accounting for the curvature of the manifold, is computed as:
\begin{equation}
\beta _{\boldsymbol{\phi }}^{\left( i \right)}=\frac{\left< \nabla _{\mathcal{R}_{\boldsymbol{\phi }}}f^{\left( i \right)},\nabla _{\mathcal{R}_{\boldsymbol{\phi }}}f^{\left( i \right)}-\nabla _{\mathcal{R}_{\boldsymbol{\phi }}}f^{\left( i-1 \right)} \right>}{ \left< \nabla _{\mathcal{R}_{\boldsymbol{\phi }}}f^{\left( i-1 \right)},\nabla _{\mathcal{R}_{\boldsymbol{\phi }}}f^{\left( i-1 \right)}  \right>},
\end{equation}
where $\left\langle \cdot, \cdot \right\rangle$ denotes the Riemannian metric (Euclidean inner product in this case). The descent direction is then updated as
\begin{equation}
\boldsymbol{\eta}^{(i)} = -\nabla _{\mathcal{R}_{\boldsymbol{\phi }}}f^{\left( i \right)}+ \beta_{\boldsymbol{\phi}}^{(i)} \mathsf{Proj}_{\boldsymbol{\phi}^{(i)}}(\boldsymbol{\eta}^{(i-1)}),
\end{equation}
where $\mathsf{Proj}_{\boldsymbol{\phi}^{(i)}}(\boldsymbol{\eta}^{(i-1)})$ is the vector transport operation that moves the previous search direction $\boldsymbol{\eta}^{(i-1)}$, also a tangent vector, from the tangent space at $\boldsymbol{\phi}^{(i-1)}$ to the tangent space at $\boldsymbol{\phi}^{(i)}$, ensuring appropriate mapping with the current tangent space and accounting for the curvature of the manifold.

\textit{Descent Direction for \(\boldsymbol{\theta}\):}
Similarly, for \(\boldsymbol{\theta}\), the update coefficient and descent direction are:
\begin{align}
&\beta _{\boldsymbol{\theta }}^{\left( i \right)}=\frac{\left< \nabla _{\mathcal{R}_{\boldsymbol{\theta }}}f^{\left( i \right)},\nabla _{\mathcal{R}_{\boldsymbol{\theta }}}f^{\left( i \right)}-\nabla _{\mathcal{R}_{\boldsymbol{\theta }}}f^{\left( i-1 \right)} \right>}{\left< \nabla _{\mathcal{R}_{\boldsymbol{\theta }}}f^{\left( i-1 \right)},\nabla _{\mathcal{R}_{\boldsymbol{\theta }}}f^{\left( i-1 \right)} \right>},
\\
&\boldsymbol{\tau}^{(i)} = -\nabla _{\mathcal{R}_{\boldsymbol{\theta }}}f^{\left( i \right)}+ \beta_{\boldsymbol{\theta}}^{(i)} \mathsf{Proj}_{\boldsymbol{\theta}^{(i)}}(\boldsymbol{\tau}^{(i-1)}).
\end{align}

\textit{Descent Direction for \(\boldsymbol{X}\):}
For the matrix variable \(\boldsymbol{X}\), the conjugate gradient update coefficient $\beta_{\boldsymbol{X}}^{(i)}$ is computed as
\begin{equation}
\beta _{\boldsymbol{X}}^{\left( i \right)}=\frac{\left< \nabla _{\mathcal{R}_{\boldsymbol{X}}}f^{\left( i \right)},\nabla _{\mathcal{R}_{\boldsymbol{X}}}f^{\left( i \right)}-\nabla _{\mathcal{R}_{\boldsymbol{X}}}f^{\left( i-1 \right)} \right>_\mathrm{F}}{\left< \nabla _{\mathcal{R}_{\boldsymbol{X}}}f^{\left( i-1 \right)},\nabla _{\mathcal{R}_{\boldsymbol{X}}}f^{\left( i-1 \right)} \right>_\mathrm{F}},
\end{equation}
where the Riemannian metric $\left\langle \cdot, \cdot \right\rangle_\mathrm{F}$ here is Frobenius inner product that multiplies the entries of two matrices and sums them up. The descent direction is then given by
\begin{equation}
\boldsymbol{\varXi}^{(i)} = -\nabla _{\mathcal{R}_{\boldsymbol{X}}}f^{\left( i \right)}+ \beta_{\boldsymbol{X}}^{(i)} \boldsymbol{\varXi}^{(i-1)},
\end{equation}
where the vector transport of the previous search direction between tangent spaces reduces to the identity mapping without modification since the multinomial manifold has a flat geometric structure with zero curvature.

\subsubsection{Retraction Operation}
After obtaining the descent directions, we update the variables by moving along these directions and subsequently map the new points back onto the manifold using retraction operations $\mathsf{Retr}(\cdot)$. This retraction ensures that the updated variables satisfy their manifold constraints, thereby allowing the optimization to proceed within the feasible region.

\textit{Update for \(\boldsymbol{\phi}\):} The updated variable \(\boldsymbol{\phi}^{(i+1)}\) is computed as:
\begin{subequations}
\begin{align}
\boldsymbol{\phi }^{\left( i+1 \right)}&=\mathsf{Retr}_{\boldsymbol{\phi }^{\left( i \right)}}( \alpha_{\boldsymbol{\phi}} ^{\left( i \right)}\boldsymbol{\eta }^{\left( i \right)} ) 
\\
&=\left[ \frac{( \boldsymbol{\phi }^{\left( i \right)}+\alpha_{\boldsymbol{\phi}}^{\left( i \right)}\boldsymbol{\eta }^{\left( i \right)}) _m}{| ( \boldsymbol{\phi }^{\left( i \right)}+\alpha_{\boldsymbol{\phi}}^{\left( i \right)}\boldsymbol{\eta }^{\left( i \right)} ) _m |} \right] ,
\end{align}
\end{subequations}
where \(\alpha_{\boldsymbol{\phi}}^{(i)}\) is the step size determined by a line search method, and the element-wise normalization ensures that \(\boldsymbol{\phi}^{(i+1)}\) lies on the complex circle manifold.

\textit{Update for \(\boldsymbol{\theta}\):} Similarly, the update for \(\boldsymbol{\theta}\) is
\begin{subequations}
\begin{align}
\boldsymbol{\theta }^{\left( i+1 \right)}&=\mathsf{Retr}_{\boldsymbol{\theta }^{\left( i \right)}}( \alpha_{\boldsymbol{\theta}}^{\left( i \right)}\boldsymbol{\tau }^{\left( i \right)} ) 
\\
&=\left[ \frac{( \boldsymbol{\theta }^{\left( i \right)}+\alpha_{\boldsymbol{\theta}}^{\left( i \right)}\boldsymbol{\tau }^{\left( i \right)})_n}{| ( \boldsymbol{\theta }^{\left( i \right)}+\alpha_{\boldsymbol{\theta}}^{\left( i \right)}\boldsymbol{\tau }^{\left( i \right)} )_n |} \right].
\end{align}
\end{subequations}

\textit{Update for \(\boldsymbol{X}\):} For \(\boldsymbol{X}\), the retraction involves projecting the updated \(\boldsymbol{X}\) back onto the multinomial manifold:
\begin{subequations}
\begin{align}
\boldsymbol{X}^{(t+1)}&=\mathsf{Retr}_{\boldsymbol{X}^{\left( i \right)}}( \alpha_{\boldsymbol{X}}^{\left( i \right)}\boldsymbol{\varXi }^{\left( i \right)} ) 
\\
& = \Pi_{\mathcal{R}_{\boldsymbol{X}}}\left( \boldsymbol{X}^t + \alpha_{\boldsymbol{X}}^{\left( i \right)} \boldsymbol{\varXi}_{\boldsymbol{X}}^t \right),
\end{align}
\end{subequations}
where $\Pi_{\mathcal{R}_{\boldsymbol{X}}}$ denotes the projection onto the multinomial manifold, which can be performed using the algorithm in [30].

\subsection{Overall Algorithm}
The proposed manifold optimization algorithm is summarized in Algorithm 1. In the outer loop, the algorithm reduces $\mu$ to tighten the LSE approximation, improving the accuracy of the surrogate objective function until the prescribed threshold is reached or numerically insolvable. In the inner loop, for a given $\mu$, Algorithm 1 iteratively updates the MIS phase shift vectors and beam pattern scheduling matrix to minimize the smoothed surrogate function, thereby maximizing the received SNR maximization across all users. We initialize the phase shifts $\boldsymbol{\phi}^{(0)}$ and $\boldsymbol{\theta}^{(0)}$, as well as the beam scheduling matrix $\boldsymbol{X}^{(0)}$, randomly within their respective manifolds that satisfy the unit-modulus and probability-simplex constraints, respectively. After reaching the stopping criteria of $\mu$, we use the thresholding technique to ensure the binary property of the beam scheduling matrix. 
\renewcommand{\algorithmicrequire}{\textbf{Input:}}
\renewcommand{\algorithmicensure}{\textbf{Output:}}
\begin{algorithm}[!t]
\caption{Manifold Optimization-Based Algorithm for (P1)}
\label{alg:general_solution}
\begin{algorithmic}[1]
\REQUIRE $N_c,N_r,M_c,M_r, \{\boldsymbol{c}_k\},\boldsymbol{\phi}, \boldsymbol{\theta}, \boldsymbol{X}$
\ENSURE $\boldsymbol{\phi}^\star, \boldsymbol{\theta}^\star,\boldsymbol{X}^\star$
\STATE Calculate the parameters related to differential position-shifting: $U_r, U_c, U, u,\{\boldsymbol{S}_u\}$, and $\{\boldsymbol{e}_u\}$.
\STATE Initialize the LSE smoothing parameter $\mu$.
\STATE Initialize $\boldsymbol{\phi}^{(0)}$, $\boldsymbol{\theta}^{(0)}$, and $\boldsymbol{X}^{(0)}$ within their respective manifolds $\mathcal{R}_{\boldsymbol{\phi}}$, $\mathcal{R}_{\boldsymbol{\theta}}$, and $\mathcal{R}_{\boldsymbol{X}}$.
\STATE Set initial descent directions $\boldsymbol{\eta}^{(0)},\boldsymbol{\tau}^{(0)}$, and $\boldsymbol{\varXi}^{(0)}$ for updating $\boldsymbol{\phi}$, $\boldsymbol{\theta}$, and $\boldsymbol{X}$, respectively.
\WHILE{stopping criteria for $\mu$ not met}
        \STATE Set iteration index $i = 0$.
        \WHILE{the norm of Riemannian gradient $\big(\lVert \nabla _{\mathcal{R}_{\boldsymbol{\phi }}}f^{(i)} \rVert _{2}^{2}+\lVert \nabla _{\mathcal{R}_{\boldsymbol{\theta }}}f^{(i)}\rVert _{2}^{2}+\lVert \nabla _{\mathcal{R}_{\boldsymbol{X}}}f^{(i)} \rVert _{\text{F}}^{2}\big)^{\frac{1}{2}}$ is above the threshold}
            \STATE Set iteration index $i = i + 1$.
            \STATE Calculate step size $\alpha_g^{(i-1)}$ using backtracking algorithms [29].
            \STATE Update $\boldsymbol{\phi}^{(i)}$ using (20a), (25), (26), and (31).
            \STATE Update $\boldsymbol{\theta}^{(i)}$ using (20b), (27), (28), and (32).
            \STATE Update $\boldsymbol{X}^{(i)}$ using (20c), (29), (30), and (33).
        \ENDWHILE
    \STATE Reduce LSE smoothing parameter $\mu=\frac{\mu}{2}$.
\ENDWHILE
\RETURN $\boldsymbol{\phi}^\star=\boldsymbol{\phi}^{(i)}$, $\boldsymbol{\theta}^\star=\boldsymbol{\theta}^{(i)}$, and $\boldsymbol{X}^\star=\boldsymbol{X}^{(i)}$ after applying binary thresholding.
\end{algorithmic}
\end{algorithm}

For a given smoothing parameter $\mu$, the algorithm uses the RCG method on compact and smooth manifolds. Under regularity conditions such as Lipschitz continuity of the gradient and appropriate step size selection during line search [29], the RCG method converges to a stationary point. Although the algorithm achieves a locally optimal solution with the smoothed surrogate objective function in each inner loop, excessively small values of $\mu$ may lead to numerical issues, hindering further refinement and terminating the process without ensuring the optimality of the original problem. Nevertheless, by carefully managing the range and update rate $\delta$ of $\mu$, the algorithm obtains feasible high-quality solutions in practice for the inherently mixed-integer, non-convex, and non-smooth problem (P1).
Let $T_{\text{inn}}$ denote the number of inner iterations and $T_{\text{out}}$ the number of outer iterations. The overall complexity of the algorithm is given by $\mathcal{O}( T_{\text{out}}T_{\text{inn}} (KUMN+KUM+( T_{\ell,3}+1) KU\log (U)+5KU+( T_{\ell,1}+4 ) M+( T_{\ell,2}+4) N ))$. Since $U$ is determined by $M$ and $N$ in (1), substituting this relationship reveals that the dominant term in the worst-case scenario scales as $\mathcal{O}(T_{\text{out}} T_{\text{inn}} KM^2 N)$.

\section{Validation Results}
In this section, we present the validation results. In Section V-A, we demonstrate the implementation and results of a proof-of-concept experiment based on the fabricated MS prototypes. Then, in Section V-B, we provide numerical simulation results to further evaluate the performance of the worst-case SNR with more flexible configurations.

\setlength{\abovecaptionskip}{6pt}
\subsection{Hardware Implementation and Experimental Results}
\begin{figure}[t]
\centering
\includegraphics[width=3.3in]{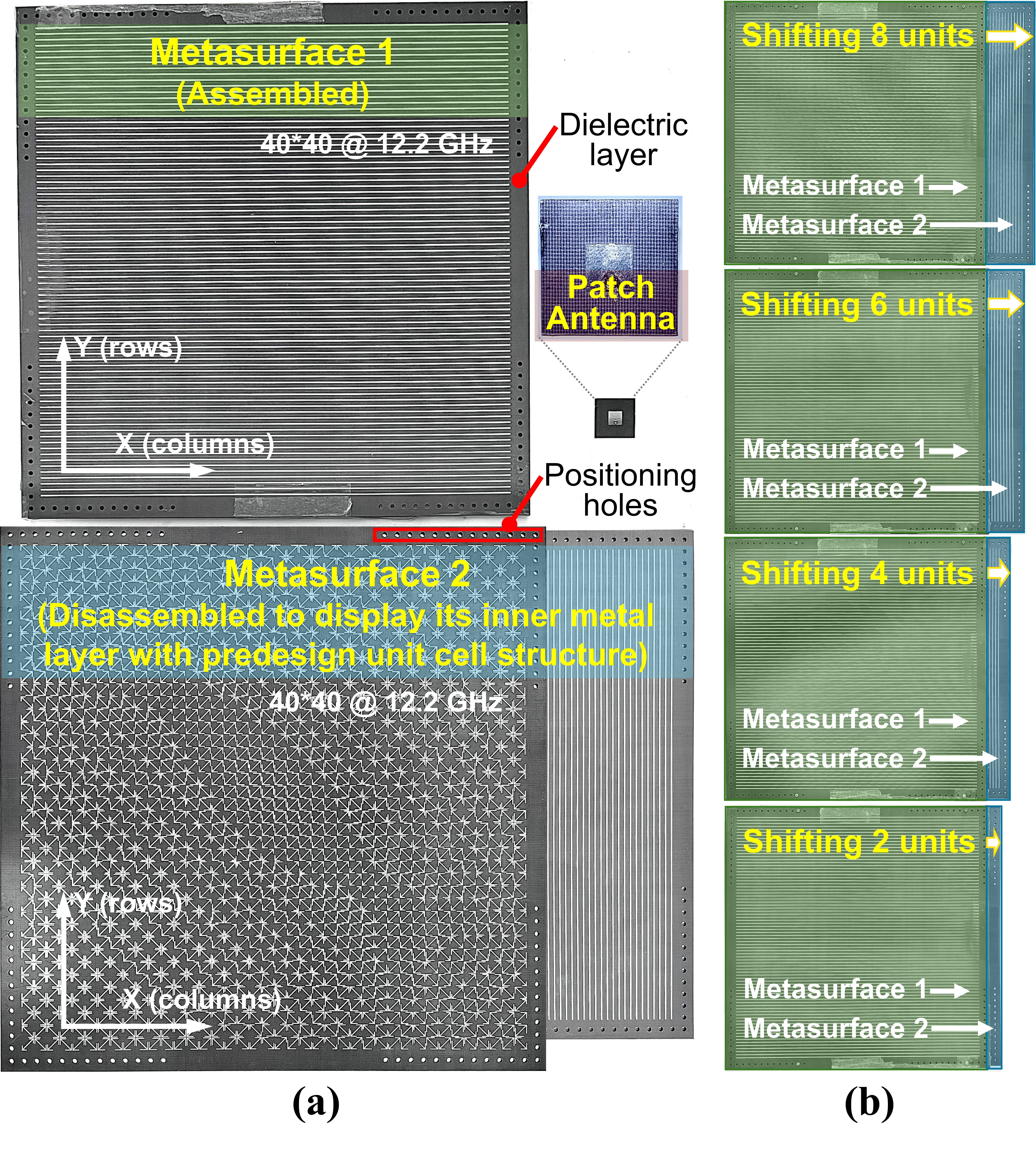}
\captionsetup{font=small}
\caption{(a) Fabricated prototypes of MS 1, MS 2, and a patch antenna; (b) closely stacked MSs with differentially shifted positions.} 
\vspace{-12pt}
\end{figure}
\begin{figure}[t]
\vspace{-4pt}
\centering
\includegraphics[width=3.3in]{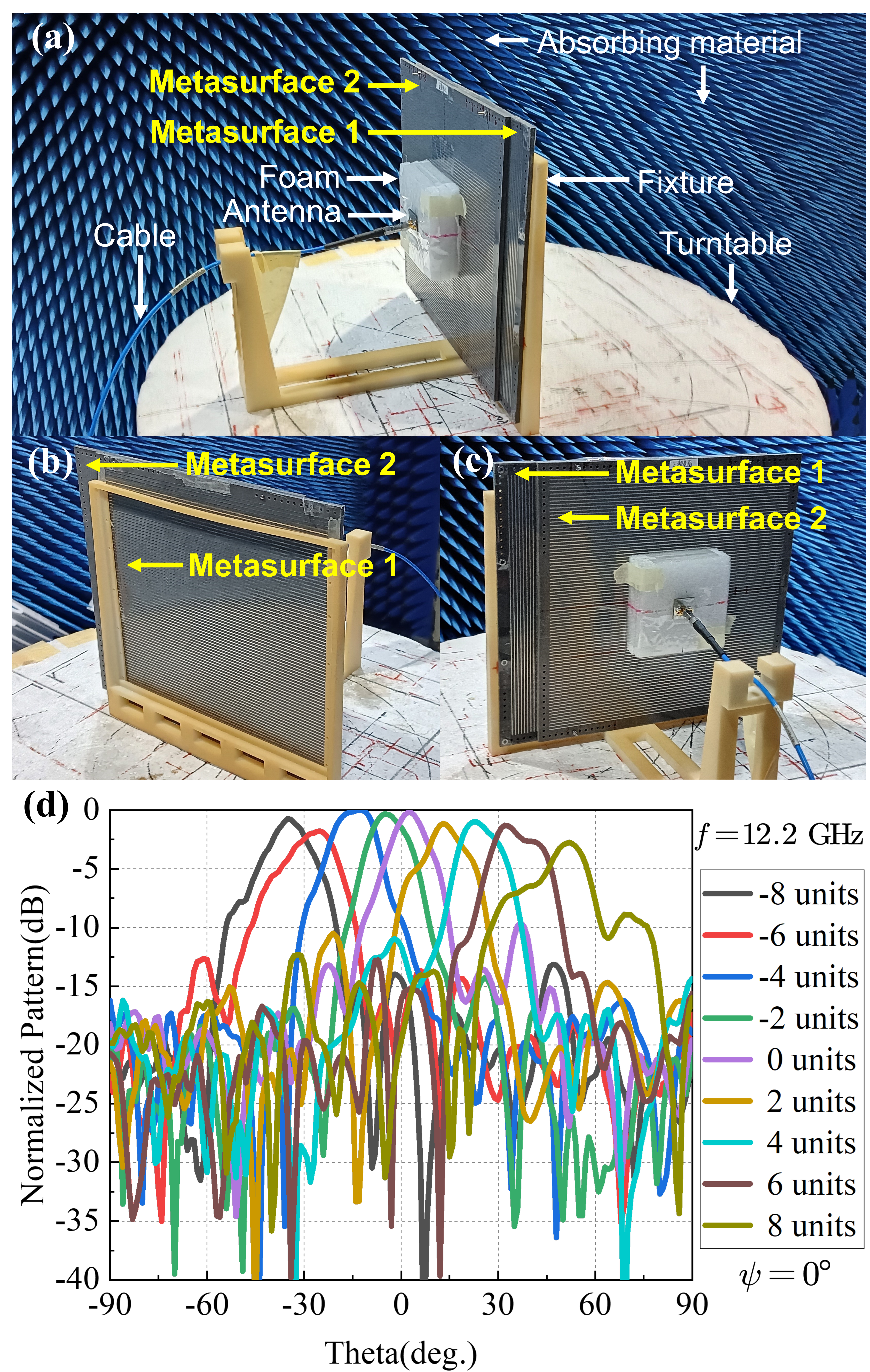}
\captionsetup{font=small}
\caption{Experimental devices in a compact range anechoic chamber for validating the 1D beam scanning: (a) Measurement setup in side view; (b) front view; (c) back view; (d) Measured normalized radiation patterns of the scanned beams on the XoZ-plane. Each beam pattern corresponds to certain shifted units in the X-direction.} 
\vspace{-12pt}
\end{figure}

\subsubsection{Prototyping and Implementation}
To validate the feasibility of beam steering enabled by the MIS architecture, we designed and manufactured prototypes of MS 1, MS 2, and a patch antenna for experimental measurements. As shown in Fig. 4(a), the fabricated transmissive MSs have a size of $240 \times 240 \, \text{mm}^2$, operate at a frequency of 12.2 GHz, and consist of $40 \times 40$ unit cells with a period of 6 mm. Specifically, each unit cell comprises two dielectric layers and three metal layers, where the top and bottom metal layers form mutually orthogonal gratings to selectively transmit and reflect waves of different polarizations. The inner metal layer has a properly designed phase distribution with rotational symmetry, as seen in the disassembled MS 2. The exposed sections of MS 1 and MS 2 exhibit orthogonal metal gratings. In addition, positioning holes with a diameter of 2 mm and a spacing of 6 mm are placed around the MSs to allow precise position tuning. Subsequently, as depicted in Fig. 4(b), two MSs are manufactured with a vertical spacing of 1 mm. MS 2 can then be shifted by 2, 4, 6, and 8 units relative to MS 1, facilitated by positioning holes.

\subsubsection{Experimental Measurements and Results}
Figs. 5(a), 5(b), and 5(c) illustrate the measurement setup from different observation views. The MS was mounted on a rotating platform (turntable) using a fixture, and the radiation patterns were measured in a compact range microwave anechoic chamber located at Shanghai Jiao Tong University. The antenna was positioned 30 mm away from MS 1, separated by foam. Absorbing materials were used to minimize reflections in the surrounding area. Fig. 5(d) shows the measured normalized radiation patterns as the position of MS 2 shifts in the X/column direction, with a displacement range from -48 mm to 48 mm in 12 mm steps, that is, -8 to 8 units in 2 unit steps. As observed in Fig. 5(d), the beam angle shifts correspondingly with the movement of MS 2, achieving the desired beam steering function with a steering angle of ±45° while maintaining a gain fluctuation of less than -3 dB. Note that given the rotational symmetry of the MSs' phase distribution, only the measured pattern results for the differential position shifting in the X/column direction are provided here, as the Y/row direction results are identical. 

\vspace{-10pt}
\subsection{Numerical simulation Results}
\vspace{-2pt}
\begin{figure}[t]
\centering
\includegraphics[width=2.5in]{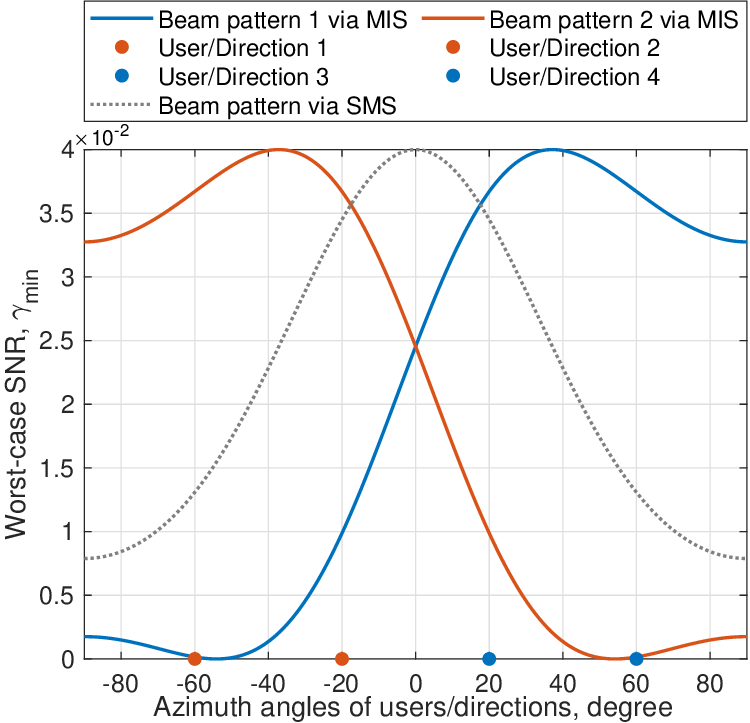}
\captionsetup{font=small}
\caption{Case study on the beam pattern of a MIS configured with $M=2\times1$ and $N=1$ versus a single-layer SMS.} 
\vspace{-6pt}
\end{figure}
\begin{figure}[t]
\centering
\includegraphics[width=2.5in]{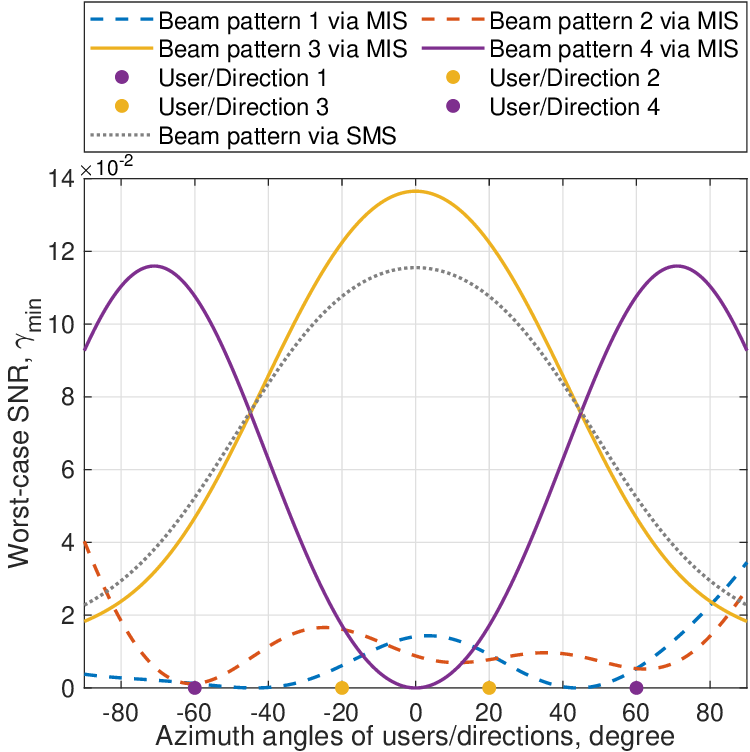}
\captionsetup{font=small}
\caption{Case study on the beam pattern of a MIS configured with $M=2\times2$ and $N=1$ versus a single-layer SMS.} 
\vspace{-15pt}
\end{figure}
This section presents numerical results to evaluate the proposed MIS beam steering schemes for coverage optimization by improving the worst-case SNR. We also draw insights into optimal MIS configuration and element allocation strategies. In the simulations, users are distributed around the MIS with elevation AoA $\psi_k=\frac{\pi}{4},\forall k$, uniformly divided azimuth AoA $\psi_k \in [-\frac{\pi}{3},\frac{\pi}{3}]$, and the same distance from MS to form the target coverage area. We fix the large-scale path loss, the transmit antenna and power at BS, and the noise power for each user, establishing a reference SNR of $\gamma_{\text{ref}}=-20$ dB, which is equivalent to set $\iota_k=0.01$ for all users in (7). 

\setlength{\abovecaptionskip}{6pt}
\begin{figure*}[t]
    \centering
    \subfloat[M=6$\times$6=36, K=8.]{
    \includegraphics[width=2.2in]{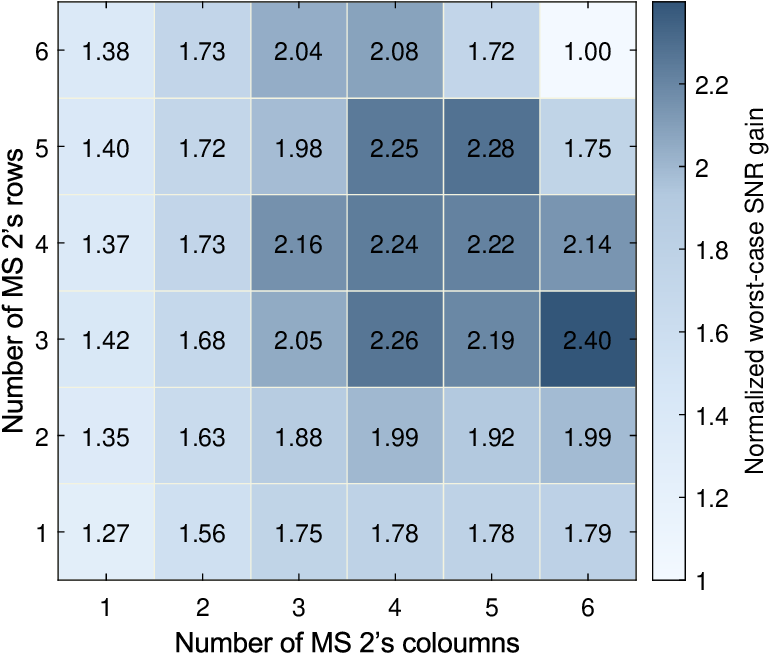}}
    \hspace{-0.05in}
    \subfloat[M=6$\times$6=36, K=16.]{
        \includegraphics[width=2.2in]{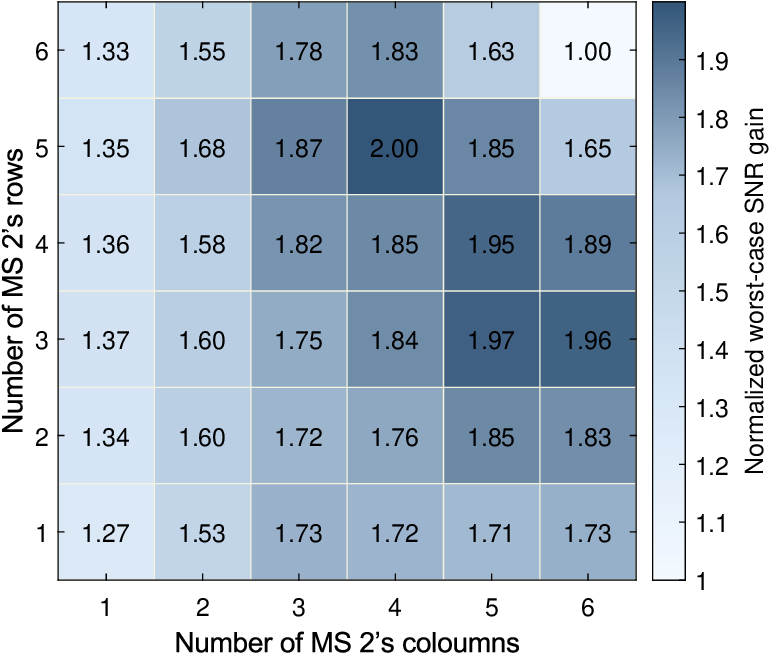}}
    \hspace{-0.05in}
    \subfloat[M=6$\times$6=36, K=32.]{
        \includegraphics[width=2.2in]{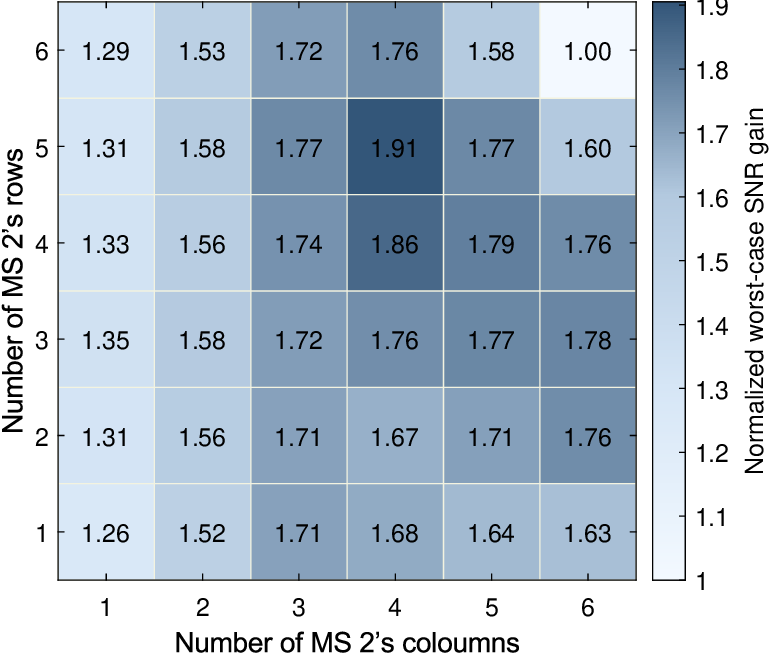}}
    \vspace{-5pt}
    \\
    \subfloat[M=8$\times$8=64, K=8.]{
        \includegraphics[width=2.2in]{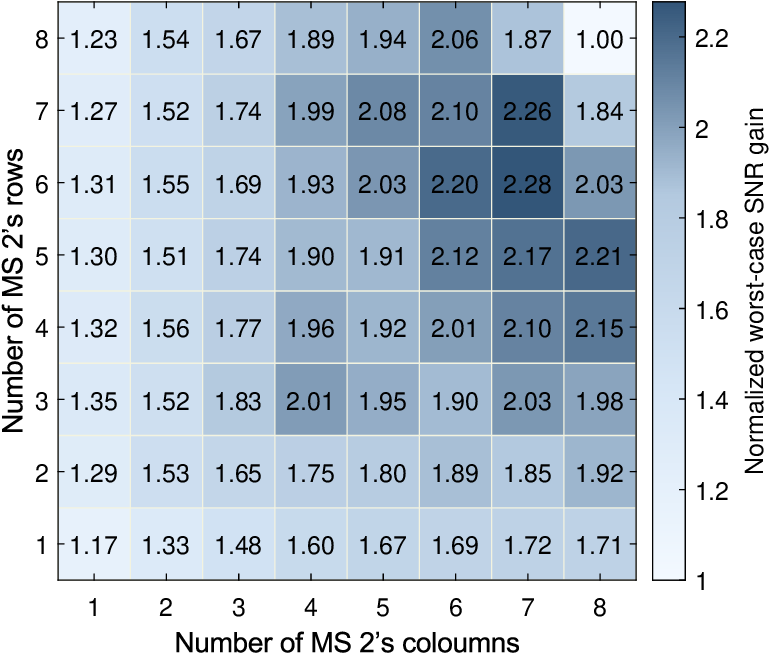}}
    \hspace{-0.05in}
    \subfloat[M=8$\times$8=64, K=16.]{
        \includegraphics[width=2.2in]{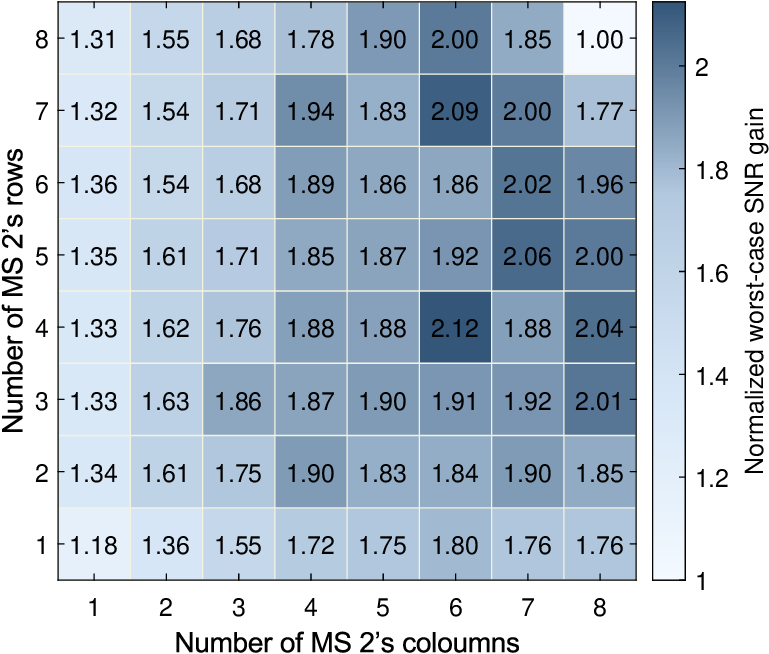}}
    \hspace{-0.05in}
    \subfloat[M=8$\times$8=64, K=32.]{
        \includegraphics[width=2.2in]{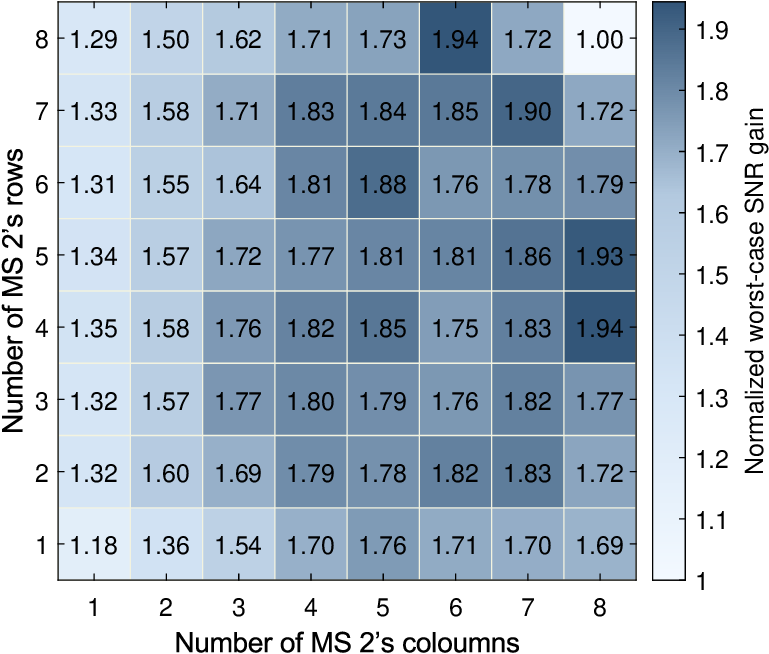}}
    \vspace{-5pt}
    \\
    \subfloat[M=10$\times$10=100, K=8.]{
        \includegraphics[width=2.2in]{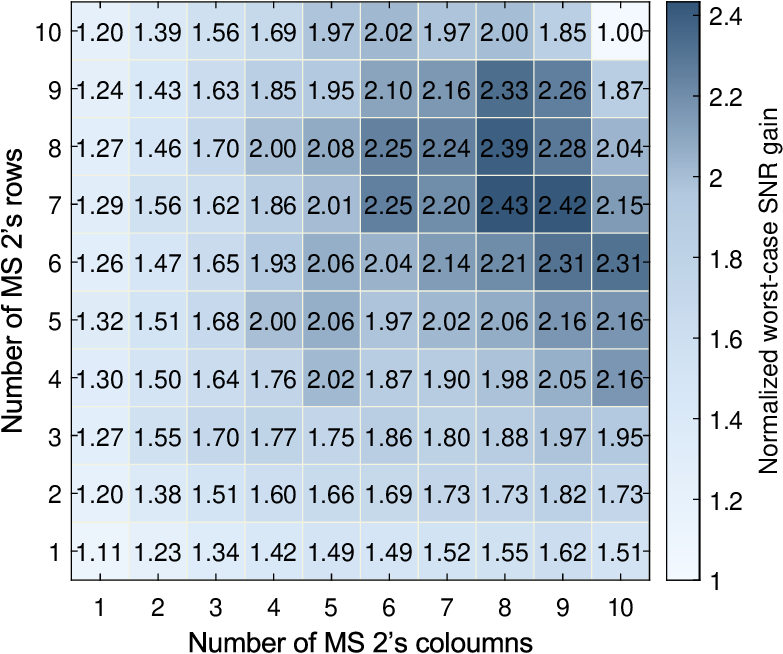}}
    \hspace{-0.05in}
    \subfloat[M=10$\times$10=100, K=16.]{
        \includegraphics[width=2.2in]{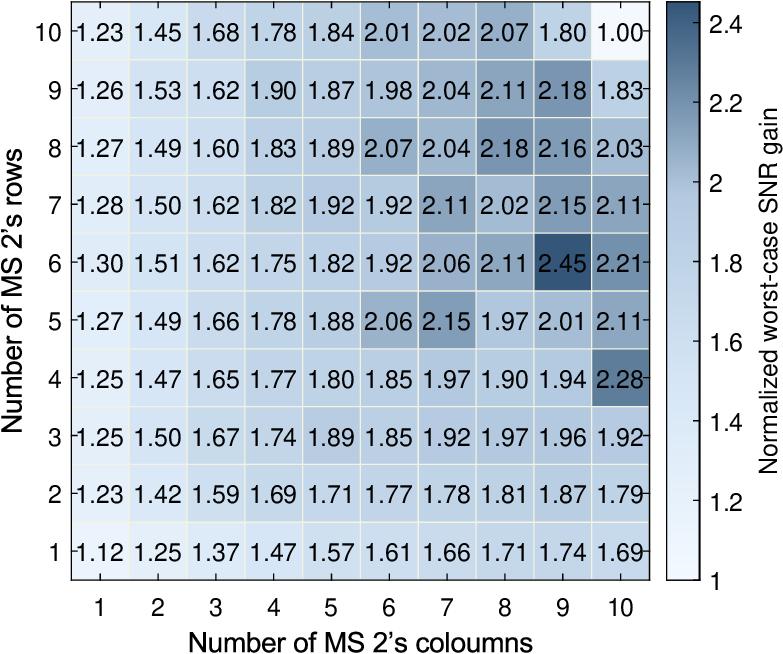}}
    \hspace{-0.05in}
    \subfloat[M=10$\times$10=100, K=32.]{
        \includegraphics[width=2.2in]{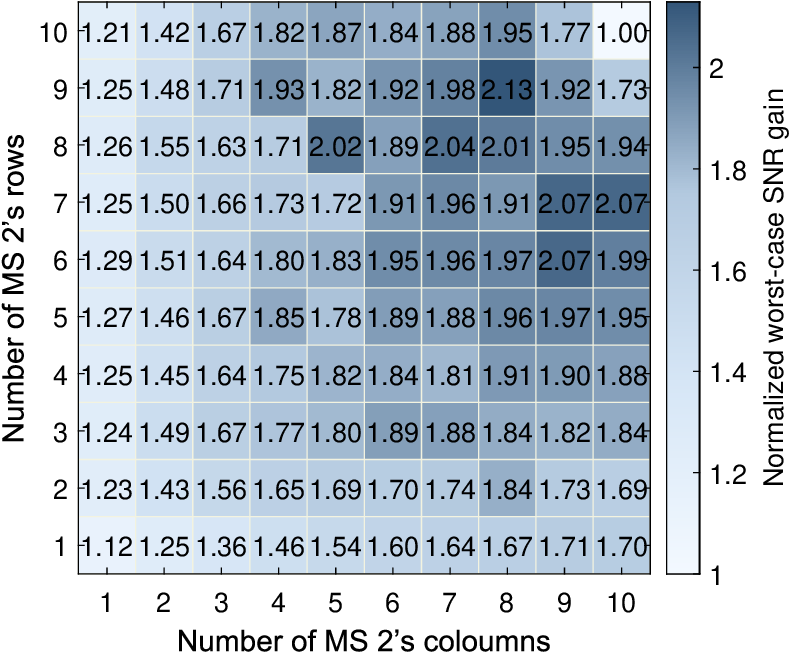}}
    \label{fig:beam_steering}
    \caption{Normalized worst-case SNR gain of the proposed MIS scheme over the single-layer SMS with varying $N_r\times N_c$, i.e., the size of MS 2, under different MS 1's size $M$ and $K$.} 
    \vspace{-12pt}
\end{figure*}

\subsubsection{Case study on the beam patterns of MISs compared with its single-layer SMS counterparts}
Figs. 6 and 7 present two case studies demonstrating MIS's ability to adjust beam patterns. In Fig. 6, the MIS is configured with $M=2\times1$ for MS 1 and $N=1$ for MS 2, allowing the generation of two distinct beam patterns to cover four users in 20° separated directions, where each user direction is represented by a specific color corresponding to their scheduled beam pattern. In contrast, the single-layer SMS generates only one beam pattern, limiting its adaptability to varying user directions. As a result, despite that both MS 1 and MS 2 have static phase shifts, the MIS achieves a significant improvement in the worst-case SNR compared to single-layer SMS, particularly enhancing the SNR for users at azimuth angles -60° and 60°. 
This performance enhancement is attributed to the dynamic beam pattern adjustment enabled by the MIS through the differential position shifting of MS 2, together with the proposed beam scheduling scheme that selects the optimal beam pattern for each user.
Fig. 7 extends this analysis by configuring the MIS with $M=2\times2$ and $N=1$, allowing the generation of four beam patterns to serve four users. Similarly to Fig. 6, MIS significantly outperforms single-layer SMS by tailoring multiple beam patterns to specific user directions. In particular, although the MIS can generate four different beam patterns, not all are utilized. Specifically, Beam patterns 1 and 2 are discarded, while Beam patterns 3 and 4 are selected to serve Users 2 and 3 and Users 1 and 4, respectively. The ability to generate and schedule multiple beam patterns is a key factor contributing to the improved performance of the MIS system.

\begin{figure}[t]
    \centering
    \subfloat[Total 64 elements.]{
        \includegraphics[width=2.5in]{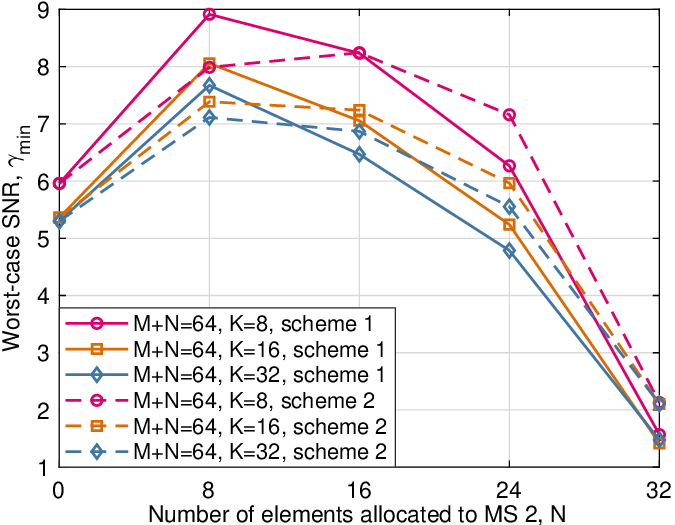}}
        \vspace{-6pt}
    \\
    \subfloat[Total 100 elements.]{
        \includegraphics[width=2.5in]{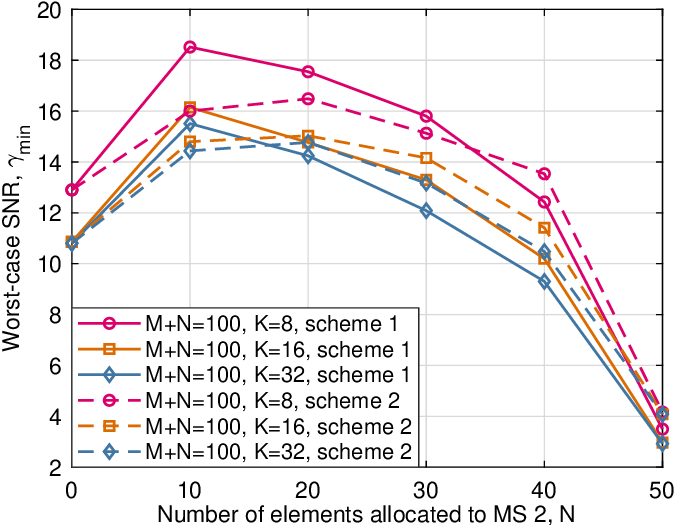}}
        \vspace{-6pt}
    \\
    \subfloat[Total 144 elements.]{
        \includegraphics[width=2.5in]{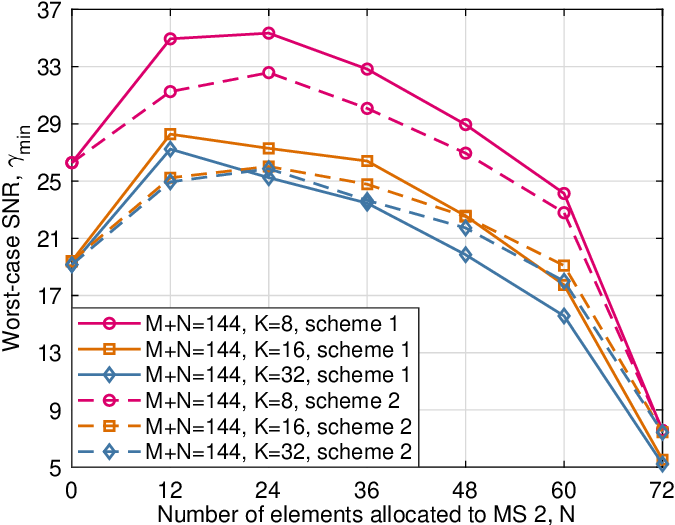}}
    \label{1fig:beam_steering}
    \captionsetup{font=small}
    \caption{Worst-case SNR performance of the MIS scheme under different strategies of element allocation between MS 1 and MS 2.} 
    \vspace{-15pt}
\end{figure}

\subsubsection{Impact of the size of MS 2 on SNR performance}
Fig. 8 presents a series of heatmaps illustrating the normalized worst-case SNR performance gain of the proposed MIS scheme over a single-layer SMS. We evaluate three MS 1 sizes with elements \(M=6 \times 6\), \(M=8 \times 8\), and \(M=10 \times 10\), which are used to cover different numbers of users (\(K = 8\), \(16\), and \(32\)), respectively. In each heatmap, the size of MS 2 (\(N_r \times N_c\)) varies from \(1 \times 1\) to \(M_r \times M_c\). When the number of MS 2's elements is equal to that of MS 1's, the configuration equivalently reduces to a single-layer SMS, serving as the baseline for normalization. The heatmaps reveal that MIS consistently outperforms single-layer SMS for each MS 1 and user configuration, with several outstanding MS 2 setups (depicted by deeper colored grids). This result highlights the importance of strategically configuring the elements of MS 2 to optimize performance. As the number of users \(K\) increases from 8 to 32, the performance gain of MIS decreases when the size of MS 1 \(M\) remains constant. This decline is attributed to the more stringent requirements of beam flattening or coverage uniformity in beam patterns when serving a larger number of users in a fair way.
Furthermore, larger sizes of MS 1 provide greater flexibility for position adjustments of MS 2, which enables more sophisticated beam patterns that better accommodate the needs of each user, especially for higher $K$. In particular, deploying a small number of MS 2 elements can yield substantial performance enhancements. Specifically, even minimal MS 2 configurations, such as a single element, achieve a gain in SNR of up to 11\% to 27\% SNR under certain configurations, revealing the substantial potential of the proposed MIS architecture for dynamic beam steering. Meanwhile, this configuration provides a cost-effective solution that balances deployment complexity with performance gains.

\subsubsection{Impact of allocation strategy of MIS elements between MS 1 and MS 2 on SNR performance}
Fig. 9 illustrates the impact on the worst-case SNR as elements are progressively allocated to MS 2 by dividing columns or rows from MS 1. The total number of elements in the MIS is fixed and assigned to MS 1 and MS 2, ensuring a fair comparison with a single-layer SMS of equivalent size. A configuration with zero elements in MS 2 reduces the system to a single-layer SMS, serving as the performance baseline. We evaluate the performance of the proposed MIS system under two element allocation schemes: Scheme 1 fixes the rows of MS 1 and the columns of MS 2 while dividing columns from MS 1 to MS 2, with the columns of MS 2 being half of the rows of MS 1 (e.g., transition from MS 1: $8 \times 8$ and MS 2: $0 \times 4$ to MS 1: $8 \times 4$ and MS 2: $8 \times 4$ when $M+N=64$); Scheme 2 fixes the columns of MS 1 and the rows of MS 2, performing operations in the row and column opposite to Scheme 1. Simulations are performed for different total elements ($M+N = 64, 100, 144$) and varying user numbers ($K = 8, 16, 32$). 
The results show that a moderate allocation of elements to MS 2 achieves the highest worst-case SNR gains, up to 47\%, compared to the single-layer SMS baseline. This significant performance gain originates from the beam steering function enabled by the MIS architecture, achieved with the same total number of elements as the traditional single-layer SMS. However, excessive allocation to MS 2 degrades performance due to insufficient resources in MS 1 for flexible beam pattern synthesis and loss of aperture gain. In the extreme case where elements are equally divided between MS 1 and MS 2, not only is the beam pattern scheduling ability nullified, but the aperture gain is also minimized. Overall, Scheme 1 exhibits similar performance to Scheme 2, achieving higher SNR peaks but decreasing faster as $N$ increases. This is because Scheme 2 offers a more balanced design, with fixed MS 1 columns that are more adaptive to the 1D user distribution. Specifically, optimal configurations occur in Scheme 1 with MS 2 allocations of $N=8$, $10$, and $12$ for total elements of $64$, $100$, and $144$, respectively. This allocation allows MS 1 to maintain its static beamforming gains while allowing MS 2 to make flexible position adjustments. Furthermore, a higher total number of elements ($M+N = 144$) leads to SNR improvements, particularly for smaller $K$, reaffirming the key insight that the optimal configuration involves allocating a small to moderate number of elements to MS 2.

\setlength{\abovecaptionskip}{3pt}
\begin{figure}[t]
\centering
\includegraphics[width=2.5in]{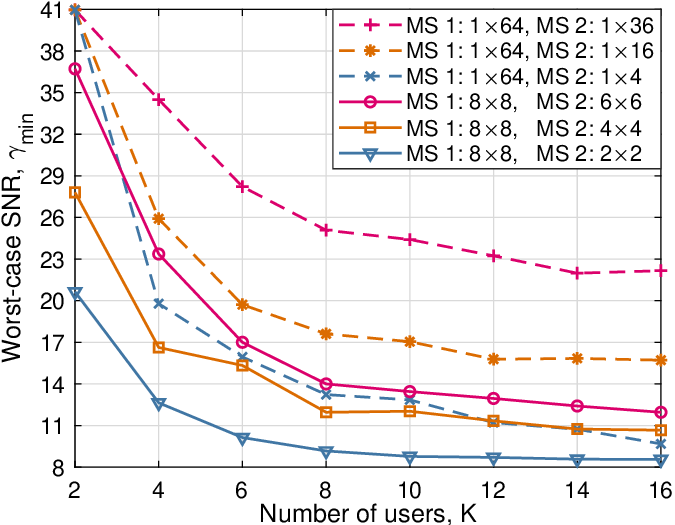}
\captionsetup{font=small}
\caption{Worst-case SINR $\gamma_{\text{min}}$ versus the number of users $K$ in the target coverage area with different 1D and 2D MIS configurations.} 
\vspace{-15pt}
\end{figure}

\subsubsection{Impact of requirements for coverage uniformity on SNR performance}
Fig. 10 illustrates the worst-case SINR $\gamma_{\text{min}}$ as a function of the number of users (\(K\)) for various 1D and 2D MIS configurations. As \(K\) increases, $\gamma_{\text{min}}$ decreases due to more stringent requirements for maintaining fairness within the target coverage area. 
The 1D MIS configuration with MS 1 comprising \(1 \times 64\) elements and MS 2 comprising \(1 \times 36\) elements achieves the highest overall SINR performance for two main reasons. First, the sufficient number of elements in MS 2 enables the synthesis of beam patterns with enough difference. Second, this configuration can generate a larger number of beam patterns (\( U = 28 \)) compared to the 2D configuration, which produces only \( U = 9 \) beam patterns with the same number of elements (MS 1: \(8 \times 8\), MS 2: \(6 \times 6\)). The increased number of beam patterns provides greater flexibility in beam scheduling, allowing better accommodation of different desired user directions. However, a 1D MIS configuration may be impractical for real-world implementation. Overall, these results emphasize two critical design considerations in system configurations: the number of beam patterns and the distinctiveness of these beam patterns.

\section{Conclusions}
In this paper, we proposed the MIS technology that enables dynamic beamforming while maintaining static phase shifts. We designed a MIS architecture comprising two closely stacked transmissive MSs: a larger fixed-position MS 1 and a smaller movable MS 2. For this architecture, we proposed the differential position shifting mechanism to synthesize distinct beam patterns, offering a cost-effective solution to achieve dynamic beamforming without element-wise phase tuning. 
Then, we modeled the interaction between two MSs and characterized a signal model for the MIS-enabled communication system. Following the modeling, we formulated an optimization problem to jointly design a set of MIS phase shifts and select shifting positions, maximizing the worst-case SNR of users in a coverage area. To tackle the intractable problem, we developed a tailored algorithm based on manifold optimization methods.
Finally, we provided extensive validation results. We first implemented a MIS prototype and conducted proof-of-concept experiments. The fabricated MIS synthesized desired beam patterns, achieving 1D beam steering with a steering angle of ±45° at 12.2 GHz and a gain fluctuation of less than -3 dB. Numerical simulations further show that deploying an MS 2 with a few elements significantly improves the worst-case SNR compared to SMSs. Furthermore, with a fixed total number of MIS elements, the strategy that allocates a moderate number of elements to MS 2 achieves optimal gains.

\ifCLASSOPTIONcaptionsoff
  \newpage
\fi




\begin{thebibliography}{99}
\bibitem{ref1}
M. Di Renzo \textit{et al.}, ``Smart radio environments empowered by reconfigurable intelligent surfaces: How it works, state of research, and the road ahead," \textit{IEEE J. Sel. Areas Commun.}, vol. 38, no. 11, pp. 2450–2525, Nov. 2020.

\bibitem{ref2}
Q. Wu and R. Zhang, ``Towards smart and reconfigurable environment: Intelligent reflecting surface aided wireless network," \textit{IEEE Commun. Mag.}, vol. 58, no. 1, pp. 106–112, Jan. 2020.

\bibitem{ref3}
Q. Wu, S. Zhang, B. Zheng, C. You, and R. Zhang, ``Intelligent reflecting surface-aided wireless communications: A tutorial," \textit{IEEE Trans. Commun.}, vol. 69, no. 5, pp. 3313–3351, May 2021.

\bibitem{ref4}
W. Mei \textit{et al.}, ``Intelligent reflecting surface-aided wireless networks: From single-reflection to multireflection design and optimization," \textit{Proc. IEEE}, vol. 110, no. 9, pp. 1380–1400, Sep. 2022.

\bibitem{ref5}
C. Pan \textit{et al.}, "An overview of signal processing techniques for RIS/IRS-aided wireless systems," \textit{IEEE J. Sel. Top. Signal Process.}, vol. 16, no. 5, pp. 883–917, Aug. 2022.

\bibitem{ref6}
Q. Wu \textit{et al.}, ``Intelligent surfaces empowered wireless network: Recent advances and the road to 6G," \textit{Proc. IEEE}, vol. 112, no. 7, pp. 724–763, Jul. 2024.

\bibitem{ref7}
S. Zeng, H. Zhang, B. Di, Z. Han, and L. Song, ``Reconfigurable intelligent surface (RIS) assisted wireless coverage extension: RIS orientation and location optimization," \textit{IEEE Commun. Lett.}, vol. 25, no. 1, pp. 269–273, Jan. 2021.

\bibitem{ref8}
G. Chen and Q. Wu, "Fundamental limits of intelligent reflecting surface aided multiuser broadcast channel," \textit{IEEE Trans. Commun.}, vol. 71, no. 10, pp. 5904–5919, Oct. 2023.

\bibitem{ref9}
Z. Zheng \textit{et al.},  ``RIS-aided hotspot capacity enhancement for multibeam satellite systems," \textit{IEEE Trans. Wireless Commun.}, vol. 23, no. 4, pp. 3648–3664, Apr. 2024.

\bibitem{ref10}
J. Dai \textit{et al.}, ``Two-timescale transmission design for RIS-aided cell-free massive MIMO systems," \textit{IEEE Trans. Wireless Commun.}, vol. 23, no. 6, pp. 6498–6517, Jun. 2024.

\bibitem{ref11}
Z. Zheng \textit{et al.}, ``Cooperative multi-satellite and multi-RIS beamforming: Enhancing LEO SatCom and mitigating LEO-GEO intersystem interference," \textit{IEEE J. Sel. Areas Commun.}, early access, 2024.

\bibitem{ref12}
Q. Wu and R. Zhang, ``Intelligent Reflecting Surface Enhanced Wireless Network via Joint Active and Passive Beamforming," \textit{IEEE Trans. Wireless Commun.}, vol. 18, no. 11, pp. 5394-5409, Nov. 2019.

\bibitem{ref13}
Z. Li \textit{et al.}, ``Transmissive RIS transceiver enabled multi-stream communication systems: Design, optimization and analysis," \textit{IEEE Internet Things J.}, early access, 2024.

\bibitem{ref14}
X. Mu \textit{et al.},  ``Simultaneously transmitting and reflecting (STAR) RIS aided wireless communications," \textit{IEEE Trans. Wireless Commun.}, vol. 21, no. 5, pp. 3083–3098, May 2022.

\bibitem{ref15}
H. Zhang and B. Di, ``Intelligent omni-surfaces: Simultaneous refraction and reflection for full-dimensional wireless communications," \textit{IEEE Commun. Surveys Tuts.}, vol. 24, no. 4, pp. 1997–2028, 4th Quart. 2022.

\bibitem{ref16}
H. Li, S. Shen, and B. Clerckx, ``Beyond diagonal reconfigurable intelligent surfaces: From transmitting and reflecting modes to single-, group-, and fully-connected architectures," \textit{IEEE Trans. Wireless Commun.}, vol. 22, no. 4, pp. 2311–2324, Apr. 2023.

\bibitem{ref17}
J. An \textit{et al.}, ``Stacked intelligent MSs for efficient holographic MIMO communications in 6G," \textit{IEEE J. Sel. Areas Commun.}, vol. 41, no. 8, pp. 2380–2396, Aug. 2023.

\bibitem{ref18}
R. Liu, Q. Wu, M. Di Renzo, and Y. Yuan, ``A path to smart radio environments: An industrial viewpoint on reconfigurable intelligent surfaces," \textit{IEEE Wireless Commun.}, vol. 29, no. 1, pp. 202–208, Feb. 2022.

\bibitem{ref19}
M.-M. Zhao, Q. Wu, M.-J. Zhao, and R. Zhang, ``Intelligent reflecting surface enhanced wireless networks: Two-timescale beamforming optimization," \textit{IEEE Trans. Wireless Commun.}, vol. 20, no. 1, pp. 2–17, Jan. 2021.

\bibitem{ref20}
H. Al-Tous and O. Tirkkonen, ``Coverage area optimized static reflecting surfaces," \textit{IEEE Trans. Wireless Commun.}, vol. 23, no. 8, pp. 9375–9388, Aug. 2024.

\bibitem{ref21}
H. Lu \textit{et al.}, ``Aerial intelligent reflecting surface: Joint placement and passive beamforming design with 3D beam flattening," \textit{IEEE Trans. Wireless Commun.}, vol. 20, no. 7, pp. 4128–4143, Jul. 2021.

\bibitem{ref22}
G. Chen, Q. Wu, C. Wu, M. Jian, Y. Chen, and W. Chen, ``Static IRS meets distributed MIMO: A new architecture for dynamic beamforming," \textit{IEEE Wireless Commun. Lett.}, vol. 12, no. 11, pp. 1866–1870, Nov. 2023.

\bibitem{ref23}
L. Zhu, W. Ma, and R. Zhang, ``Movable antennas for wireless communication: Opportunities and challenges," \textit{IEEE Commun. Mag.}, early access, Oct. 16, 2023, doi: 10.1109/mcom.001.2300212.

\bibitem{ref24}
K.-K. Wong, A. Shojaeifard, K.-F. Tong, and Y. Zhang, ``Fluid antenna systems," \textit{IEEE Trans. Wireless Commun.}, vol. 20, no. 3, pp. 1950–1962, Mar. 2021.

\bibitem{ref25}
L. Zhu, W. Ma, and R. Zhang, ``Movable-antenna array enhanced beamforming: Achieving full array gain with null steering," \textit{IEEE Commun. Lett.}, early access, Oct. 11, 2023, doi: 10.1109/lcomm.2023.3323656.

\bibitem{ref26}
Z. Zheng, Q. Wu, W. Chen, and G. Hu, ``Two-timescale design for movable antennas enabled-multiuser MIMO systems," arXiv:2410.05912, 2024. [Online]. Available: http://arxiv.org/abs/2410.05912

\bibitem{ref27}
H. Wang \textit{et al.}, ``Throughput maximization for movable antenna systems with movement delay consideration," arXiv:2411.13785, 2024. [Online]. Available: https://arxiv.org/abs/2411.13785

\bibitem{ref28}
E. Polak, J. O. Royset, and R. S. Womersley, ``Algorithms with adaptive smoothing for finite minimax problems," \textit{J. Optim. Theory Appl.}, vol. 119, no. 3, pp. 459–484, Dec. 2003.

\bibitem{ref29}
N. Boumal \textit{et al.}, ``Manopt, a Matlab toolbox for optimization on manifolds," \textit{J. Mach. Learn. Res.}, vol. 15, no. 42, pp. 1455–1459, 2014. [Online]. Available: https://www.manopt.org

\bibitem{ref30}
J. Duchi, S. Shalev-Shwartz, Y. Singer, and T. Chandra, ``Efficient projections onto the $\ell_1$-ball for learning in high dimensions," in \textit{Proc. 25th Int. Conf. Mach. Learn. (ICML)}, 2008, pp. 272–279.




















\end{thebibliography}
%

\end{document}